\documentclass[twocolumn,final]{revtex4}
\usepackage{amsmath,amssymb}
\usepackage{graphicx,pslatex}
\usepackage{braket}

\usepackage{color}

\usepackage{subfigure}

\begin{document}

\title{{\bf The lattice Landau gauge gluon propagator: lattice spacing and volume dependence}}
\author{Orlando Oliveira}
\email{orlando@teor.fis.uc.pt}
\affiliation{Centro de F\'{i}sica Computacional, Departamento de F\'{i}sica, Universidade de Coimbra, 3004-516
Coimbra, Portugal}

\author{Paulo J.~Silva}
\email{psilva@teor.fis.uc.pt}
\affiliation{Centro de F\'{i}sica Computacional, Departamento de F\'{i}sica, Universidade de Coimbra, 3004-516
Coimbra, Portugal}

\begin{abstract}
The interplay between the finite volume and finite lattice spacing is investigated using lattice QCD simulations
to compute the Landau gauge gluon propagator. 
Comparing several ensembles with different lattice spacings and physical volumes, we conclude that the 
dominant effects, in the infrared region, are associated with the use of a finite lattice spacing. 
The simulations show that decreasing the lattice spacing, while keeping the same physical volume, leads
to an enhancement of the infrared gluon propagator. In this sense, the data from 
$\beta=5.7$ simulations, which uses an $a \approx 0.18$ fm, provides a lower bound for the infinite volume propagator. 
\end{abstract}
\maketitle

\section{Introduction and Motivation}

In the past years the Landau gauge gluon and ghost propagators have been studied using lattice QCD methods
in pure Yang-Mills theories -- see, for example, \cite{ovr, Silva2004,Leinweber1999, Silva2006, largevolume, cucc1, cucc2, cucc3, Bonnet01, bounds, ratios, qcdtnt11, maas, boucaud, furui, Sternbeck2012} 
and references therein. The main goal is to be able to compute the infrared propagators having control
on the finite size and finite volume effects.

The lattice simulations were performed using large physical volumes, let us say above (3 fm)$^4$, both for the
SU(2) and SU(3) gauge groups. The largest volumes simulated so far being $\sim (27 \mbox{ fm})^4$ for the 
SU(2) gauge group \cite{cucc2} and $\sim (17 \mbox{ fm})^4$ for SU(3) \cite{largevolume}. To achieve such large 
volumes, the simulations 
use a relatively large lattice spacing, $\sim 0.2$ fm to be compared with the typical 
non-perturbative scale in QCD of $\sim 1$ fm. It is well known that the infrared propagator decreases as
the physical volume is increased and the main concern about the recent simulations has been the
control of the finite volume effects. For such large volumes, which are much larger than typical nuclear sizes,
finite volume effects should be negligible or smaller than other possible systematics. Less
studied is the interplay between finite lattice spacing and finite volume effects. Indeed, given the large lattice
spacing used to simulate such large physical volumes, the interplay between the two effects can introduce
a bias on the propagators.

The aim of the present work is to investigate the interplay between finite lattice spacing and finite volume effects
for the pure SU(3) Yang-Mills theory and, in particular, for the gluon propagator. 
In principle our conclusions should also be valid for the SU(2) simulations. Having control of all possible effects
is crucial to compare lattice results with the outcome of other non-perturbative techniques. Furthermore, being
able to remove the finite size and volume effects, the propagators can be used to model more reliably 
non-perturbative physics.

In order to understand how the finite lattice spacing and the finite lattice volume change the propagator,
we perform various simulations with several lattice spacings and physical volumes and compare the 
propagators. The simulations show, once more, that the infrared propagator decreases when the
physical volume of the simulation is increased. However, they also show that the finite lattice spacing 
effects have a much larger impact on the infrared propagator. The propagators are qualitatively unchanged
but we observe that a large lattice spacing underestimates the gluon propagator in the infrared region.

The simulations also show that we have now good control on the Landau gauge gluon propagator for
momenta above 900 MeV. Furthermore, the extrapolations to the infinite volume discussed here suggest that
we are able to provide a gluon propagator free of finite size effects for momenta above 400 MeV.

The paper is organized as follows. In section \ref{Section:Renor} we discuss the lattice setup and renormalization
procedure. In section \ref{Volume}, profiting from having various ensembles with various lattice spacings and
volumes, the extrapolation of the zero momentum gluon propagator to the infinite volume limit is investigated.
In section \ref{spacing} the lattice spacing effects on the infrared gluon propagator are investigated.
In section \ref{glue_D0} we model the propagator in order to be able to
 provide an extrapolation to the infinite volume.
Finally, in section \ref{resultados} we resume and conclude.

\section{Lattice Setup and Renormalization Procedure \label{Section:Renor}}

In the current work we report on simulations for the pure gauge SU(3) Yang-Mills theory using the
Wilson action at several $\beta$ values and for different physical volumes. 
Table \ref{tab:latsetup} lists the various ensembles used to compute the gluon propagator.

The gauge configurations were generated with the MILC code \cite{milc} using a combined Monte Carlo sweep 
of seven overrelaxation updates with four heat bath updates. For each ensemble the autocorrelation time
was measured for the plaquette. For all the ensembles, the autocorrelation time was below 10 combined
Monte Carlo sweep. The gluon propagator was measured by a separation of ten autocorrelation
times, with a minimum separation of 100 combined sweeps. For thermalization, we disregarded the first
500 combined Monte Carlo sweeps. For the conversion to physical units we use the lattice spacing as measured
from the string tension \cite{Bali1993}.

The gauge links were rotated to the minimal Landau gauge by minimizing the function
\begin{equation}
  F[g] = \frac{1}{V \, N_d N_c} \sum_{x,\mu} \, \mbox{Re Tr} \left[ U^g_\mu (x) \right]
  \label{Eq:Fg}
\end{equation}  
where $V$ is the number of lattice points, $N_d = 4$ the number of space-time dimensions, $N_c = 3$ the number
of colors and $U^g_\mu (x)$ the gauge transformed link.
 For the minimization of $F[g]$
  over the gauge orbits we used an overrelaxation algorithm \cite{ovr}, except for 
  the $\beta = 5.7$ and $V = 18^4$, 
  $26^4$, $36^4$ and $\beta = 6.0$ for
$V = 32^4$ ensembles, where the configurations were rotated to the Landau 
gauge using a Fourier accelerated steepest descent method \cite{fasd}.
The quality of the gauge fixing was monitored by looking
at $\theta$, the lattice version of $\partial \cdot A ^a$ averaged over all lattice points and all colors; 
see, for example, \cite{Silva2004} for details and definitions. The minimization was stopped when
$\theta < 10^{-13}$.

For the computation of the gluon propagator we follow the definitions given in \cite{Silva2004}. Further,
in the following we use the tree level improved momentum definition
\begin{equation}
q_\mu = \frac{2}{a} \, \sin \left( \frac{\pi n}{L_\mu} \right) \, \qquad n = 0, 1, \dots , \frac{L_\mu}{2} \, , 
\label{Eq:momdef}
\end{equation}
where $a$ stands for the lattice spacing and $L_\mu$ the number of lattice points in direction $\mu$.
The statistical errors on the propagators were evaluated with the jackknife method.

\begin{table}[t]
   \centering
   \begin{tabular}{l@{\hspace{0.5cm}} l @{\hspace{0.5cm}} l @{\hspace{0.5cm}} l @{\hspace{0.5cm}} l @{\hspace{0.5cm}}r} 
      \toprule
      $\beta$    &    $a$ ($fm$)   &  $1/a$ (GeV) & L   & $La$ ($fm$)  &  \# Conf \\
      \hline
      5.7           &   0.1838(11) & 1.0734(63)   &  44  &  8.087     &  55 \\
      5.7           &   0.1838(11) & 1.0734(63)   &  36  &  6.617     &  100 \\
      5.7           &   0.1838(11) & 1.0734(63)   &  26  &  4.780     &  132 \\
      5.7           &   0.1838(11) & 1.0734(63)   &  18  &  3.308     &  149 \\
      \hline
      6.0           &    0.1016(25)  & 1.943(47)  &  80  &  8.128     &  55 \\
      6.0           &    0.1016(25)  & 1.943(47)  &  64  &  6.502     &  121 \\
      6.0           &    0.1016(25)  & 1.943(47)  &  48  &  4.877     &  104 \\
      6.0           &    0.1016(25)  & 1.943(47)  &  32  &  3.251     &  100 \\
      \hline
      6.2           &    0.07261(85)  & 2.718(32) &  80  &  5.808     &  70 \\
      6.2           &    0.07261(85)  & 2.718(32) &  64  &  4.646     &  99 \\
      6.2           &    0.07261(85)  & 2.718(32) &  48  &  3.485     &  87 \\
      \hline
      6.4           &    0.05449(54)  & 3.621(36) &  80  &  4.359     &  52 \\
      \toprule
   \end{tabular}
   \caption{Lattice setup.}
   \label{tab:latsetup}
\end{table}

\begin{table}[t]
   \centering
   \begin{tabular}{l@{\hspace{0.5cm}} l @{\hspace{0.5cm}} l @{\hspace{0.5cm}} l @{\hspace{0.5cm}} l} 
      \toprule
      $\beta$    &    $a$ ($fm$)   &  $1/a$ (GeV) & L   & $La$ ($fm$)  \\
      \hline
      5.7           &   0.1838(11) & 1.0734(63)    &  64  &  11.763     \\
      5.7           &   0.1838(11) & 1.0734(63)    &  72  &  13.234     \\
      5.7           &   0.1838(11) & 1.0734(63)    &  80  &  14.704     \\
      5.7           &   0.1838(11) & 1.0734(63)    &  88  &  16.174    \\
      5.7           &   0.1838(11) & 1.0734(63)    &  96  &  17.645    \\
      \toprule
   \end{tabular}
   \caption{Lattice setup for the data taken from reference \cite{largevolume}. Note that the  
                 Berlin-Moscow-Adelaide data was rescaled to be coherent with our lattice spacing definitions -- see
                 text for details.}
   \label{tab:latsetuplarge}
\end{table}

The propagators computed with the ensembles listed in Tab. \ref{tab:latsetup} will be compared with the large 
volume simulations performed by the Berlin-Moscow-Adelaide group \cite{largevolume}.
We would like to call the reader attention that our simulations and those performed in
\cite{largevolume} use the same lattice action and the same definitions for $D(q^2)$ and the lattice momenta
(\ref{Eq:momdef}).
Table \ref{tab:latsetuplarge} summarizes the data generated by this collaboration. Note that both sets 
in Tab. \ref{tab:latsetup} and \ref{tab:latsetuplarge} use the Wilson action. Note also, that the
data in Tab. \ref{tab:latsetuplarge} reaches a physical volume which is about twice the largest volume reported 
in
Tab. \ref{tab:latsetup}. Moreover, the data reported in \cite{largevolume} use the lattice spacing computed from
$r_0$ \cite{Necco2002}, while we use the lattice spacing measured from the string tension \cite{Bali1993}.
The values are not compatible within
errors and, therefore,  to compare the Berlin-Moscow-Adelaide results with ours, we have rescaled their data 
accordingly.
In the following, we will compare only renormalised data. To correct the data of \cite{largevolume}
for to the different choices of lattice spacing, it is sufficient to rescale the momenta. 
The renormalisation procedure, as described below, takes care of setting $D(q^2)$
properly.

The simulations using the ensembles listed in table \ref{tab:latsetup} and \ref{tab:latsetuplarge} 
are associated with different lattice spacings. In order to be able to compare the various sets, 
one needs to renormalize the propagator data. As described in the next section, the renormalization
is performed at the scale $\mu = 4$ GeV and, therefore, any differences should show up in the infrared
region. 

The gluon data shown below has momentum cuts to reduce the lattice artifacts. For momentum above
1 GeV only the momentum which verify the conical plus cylindrical cuts
 \cite{Leinweber1999} are included in our analysis.
 For momentum below 1 GeV, we have considered all the lattice data.

\subsection{Renormalization \label{renormalization_1}}

In order to renormalize the lattice data the bare gluon propagator is fitted to the 1-loop inspired functional form
\begin{equation}
  D(q^2) ~ = ~ Z \, \frac{\left[ \ln \left(\frac{q^2}{\Lambda^2}\right) \right]^{-\gamma}}{q^2} \, ,
  \label{uvfit}
\end{equation}
where $\gamma = 13/22$ is the anomalous gluon dimension for pure SU(3) Yang-Mills theory, 
in the momentum range $[q_{min} , q_{max}]$. The renormalized gluon propagator is given by
\begin{equation}
   D(q^2) = Z_R \, D_{Lat} (q^2) \, ,
   \label{Eq:DRDLAT}
\end{equation}   
where $Z_R$ is the renormalization constant and $D_{Lat} (q^2)$ the bare lattice gluon propagator. 
The renormalization constant $Z_R$ is chosen such that
\begin{equation}
   \left. D(q^2) \frac{}{} \right|_{q^2 = \mu^2} = \frac{1}{\mu^2} \, 
   \label{Eq:Rpoint}
\end{equation} 
and here we will use $\mu = 4$ GeV. This choice for the renormalization scale will allow a direct
comparison with the extrapolation to the infinite volume of $D(0)$ performed in \cite{Bonnet01}.

\begin{figure*}[t] 
   \centering
   \subfigure[$\beta = 5.7$ ]{ \includegraphics[scale=0.35]{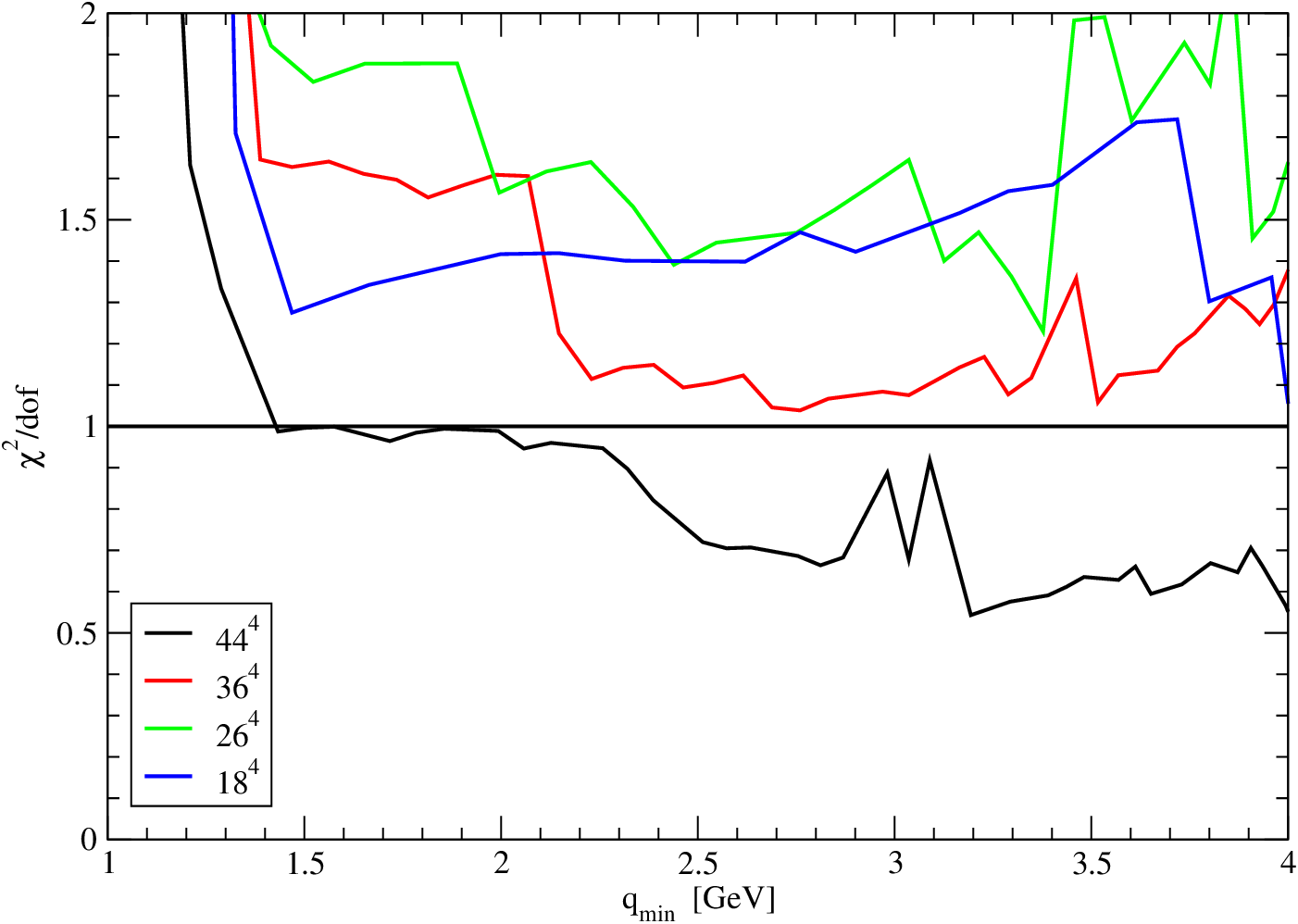} } \qquad
   \subfigure[$\beta = 6.0$ ]{ \includegraphics[scale=0.35]{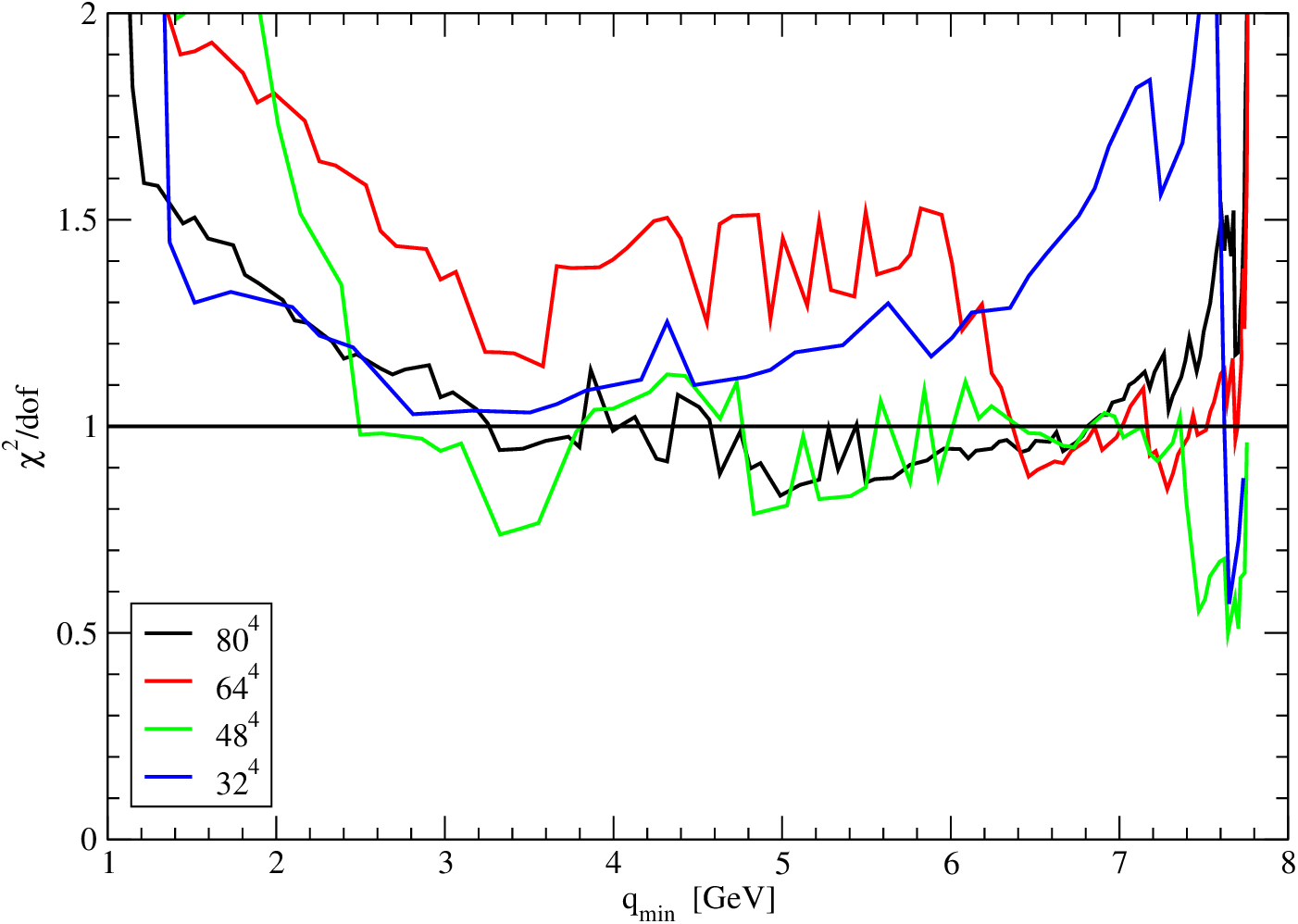} }
   
\vspace{0.65cm}
   \subfigure[$\beta = 6.2$]{ \includegraphics[scale=0.35]{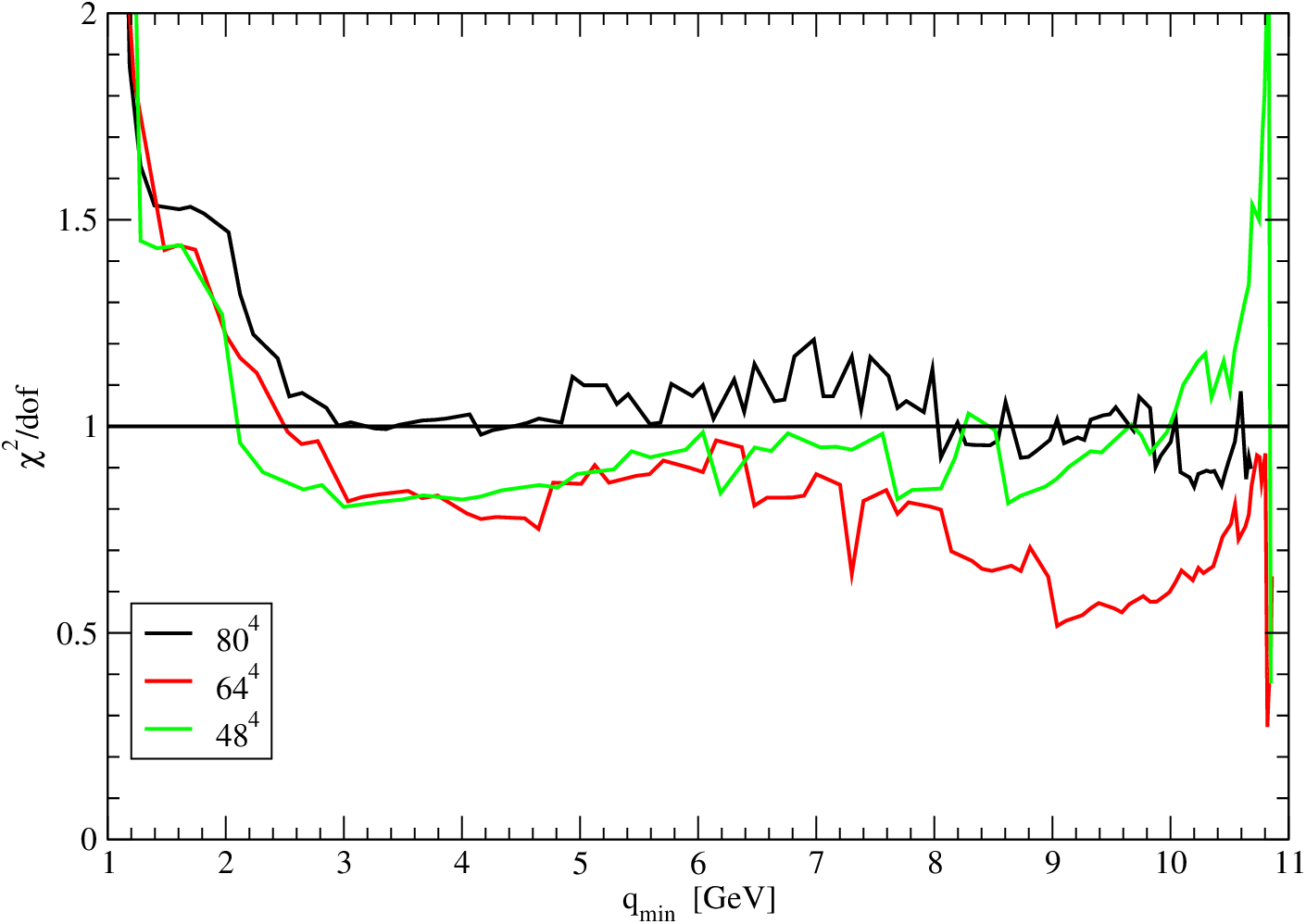} } \qquad
   \subfigure[$\beta = 6.4$]{ \includegraphics[scale=0.35]{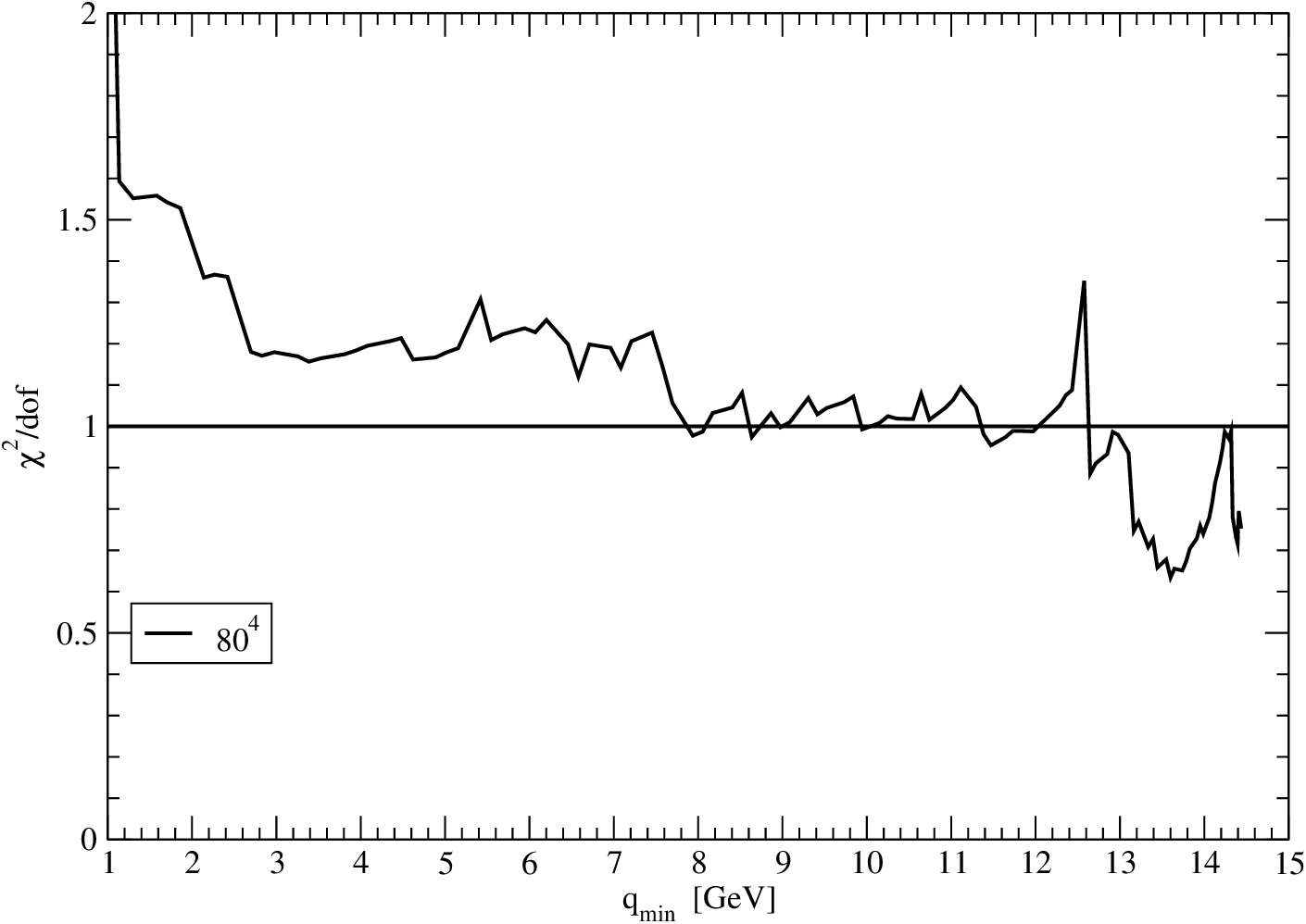} }
  \caption{$\chi^2/d.o.f.$ for the fits to Eq. (\ref{uvfit}) as a function of $q_{min}$.}
   \label{fig:chi2_UV}
\end{figure*}

In what concerns the fitting range, for the highest momentum $q_{max}$, we will use the highest 
momentum achieved in the simulation, namely $q_{max} = $ 4.3 GeV for the $\beta = 5.7$ ensembles, 
$q_{max} = $ 7.8 GeV for the $\beta = 6.0$, $q_{max} = $ 10.9 GeV for $\beta = 6.2$ and 
$q_{max} = $ 14.5 GeV for $\beta = 6.4$ data. 
The smallest fitting momentum $q_{min}$ is chosen as the smallest momentum where 
$\chi^2/d.o.f. \approx 1$ or is closer to one. Given that the renormalization scale used throughout this
work is $\mu = 4$ GeV, we also require $q_{min}$ to be smaller than 4 GeV.
The $\chi^2/d.o.f.$ as function of $q_{min}$ can be seen in Fig.
\ref{fig:chi2_UV}. 

The fitting range for the various ensembles considered in the present work are
\begin{widetext}
\begin{displaymath}
\begin{array}{l@{\hspace{0.4cm}}llr@{\hspace{0.7cm}}l@{\hspace{0.4cm}}llr
                                                       @{\hspace{0.7cm}}l@{\hspace{0.4cm}}llr
                                                       @{\hspace{0.7cm}}l@{\hspace{0.4cm}}llr}
                      &  L & q_{min} & q_{max}  &                     &  L & q_{min} & q_{max} 
                      &                     &  L & q_{min} & q_{max}
                      &                     &  L & q_{min} & q_{max} \\
  \beta = 5.7  & 44  & 1.43 & 4.29            &  \beta = 6.0  & 80 & 3.25 & 7.77       
                                                                &  \beta = 6.2  & 80 & 2.95 &10.87  
                                                                &  \beta = 6.4  & 80 & 2.70 & 14.50     \\
                     & 36  &  2.23 & 4.29           &                     & 64 & 3.24     & 7.77       
                                                                &                     & 64  & 2.52    & 10.87  
                                                                &                     &       &            &   \\
                     & 26  &  2.44 & 4.29           &                     & 48 & 2.50     & 7.77       
                                                                &                     & 48 & 2.12     & 10.87 
                                                                &                     &      &             &   
                                                                                                \\
                     & 18  & 1.47  & 4.29           &                     & 32 & 2.81     & 7.77
                                                                &                     &      &             &
                                                                &                     &      &             &   
\end{array}
\end{displaymath}
\end{widetext}
with all momenta given in GeV. Proceeding in the same way, the $q_{min}$ used to compute $Z_R$
for the Berlin-Moscow-Adelaide propagator data is 3.38 GeV for the $64^4$ lattice data, 
3.23 GeV for $72^4$, 2.76 GeV for $80^4$, 3.19 GeV for $88^4$ 
and 1.08 GeV for $96^4$ gluon propagator data. 

\begin{figure}[t] 
   \centering
   \includegraphics[scale=0.35]{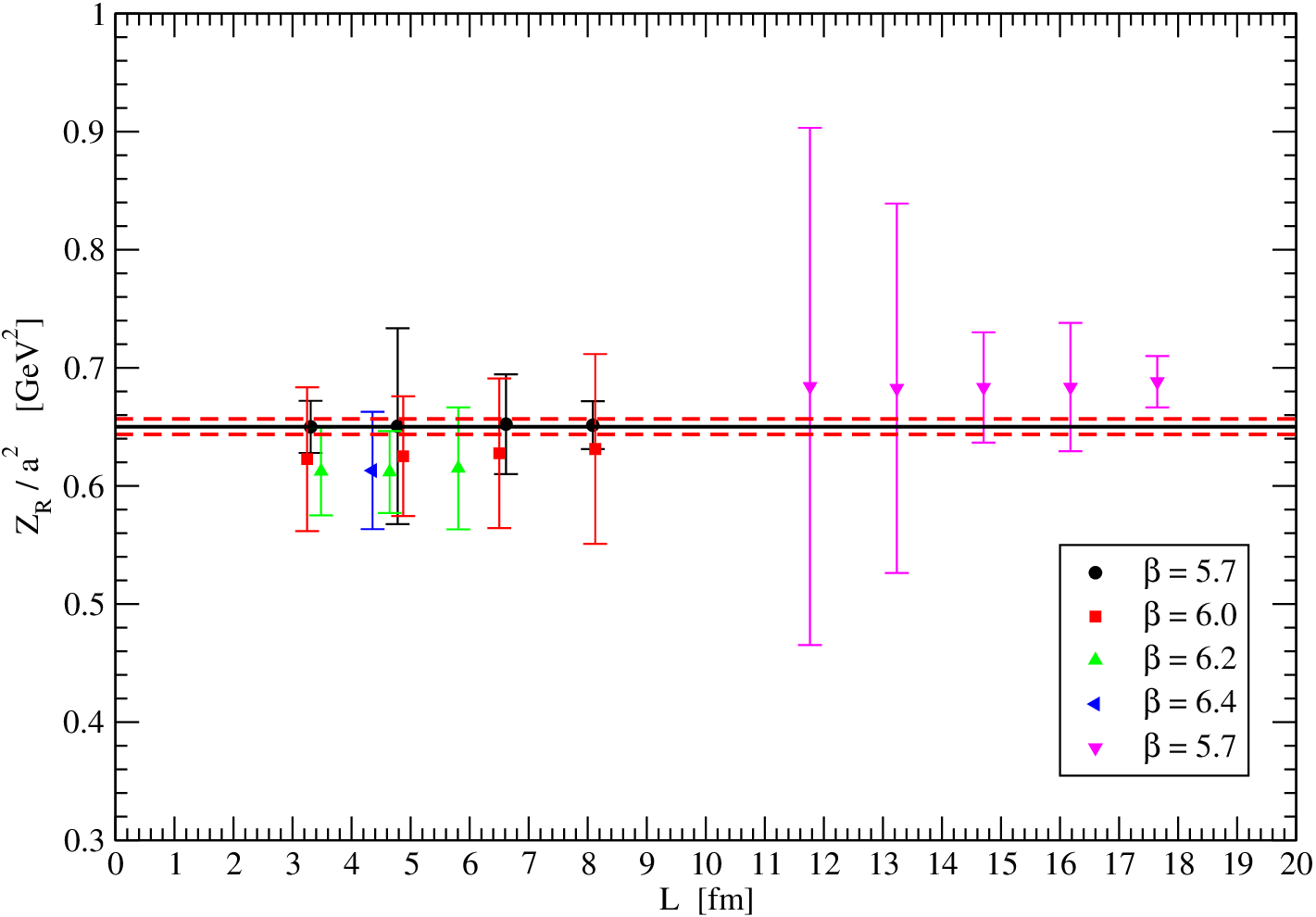} 
   \caption{Renormalization constant for the different simulations. The data is volume independent and
                 a fit to a constant gives $Z_R / a^2 = 0.6514(64)$ GeV$^2$. The continuum line is the central value
                 of the fitted data and the dashed lines are the one standard deviation values. }
   \label{fig:ZR_over_a2}
\end{figure}

The values of $Z_R / a^2$ are reported in Fig. \ref{fig:ZR_over_a2}. 
It follows  that $Z_R / a^2$ is volume independent, at least within the statistical precision
of the simulations reported here. Fitting the plotted data to a constant gives
\begin{equation}
   \frac{Z_R}{a^2} = 0.6501(65) \mbox{ GeV}^2 
   \label{Eq:fitted_ZR}
\end{equation}
with a $\chi^2/d.o.f. = 0.50$. The errors were computed assuming Gaussian error propagation. 

\begin{figure*}[t] 
   \centering
   \subfigure{ \includegraphics[scale=0.35]{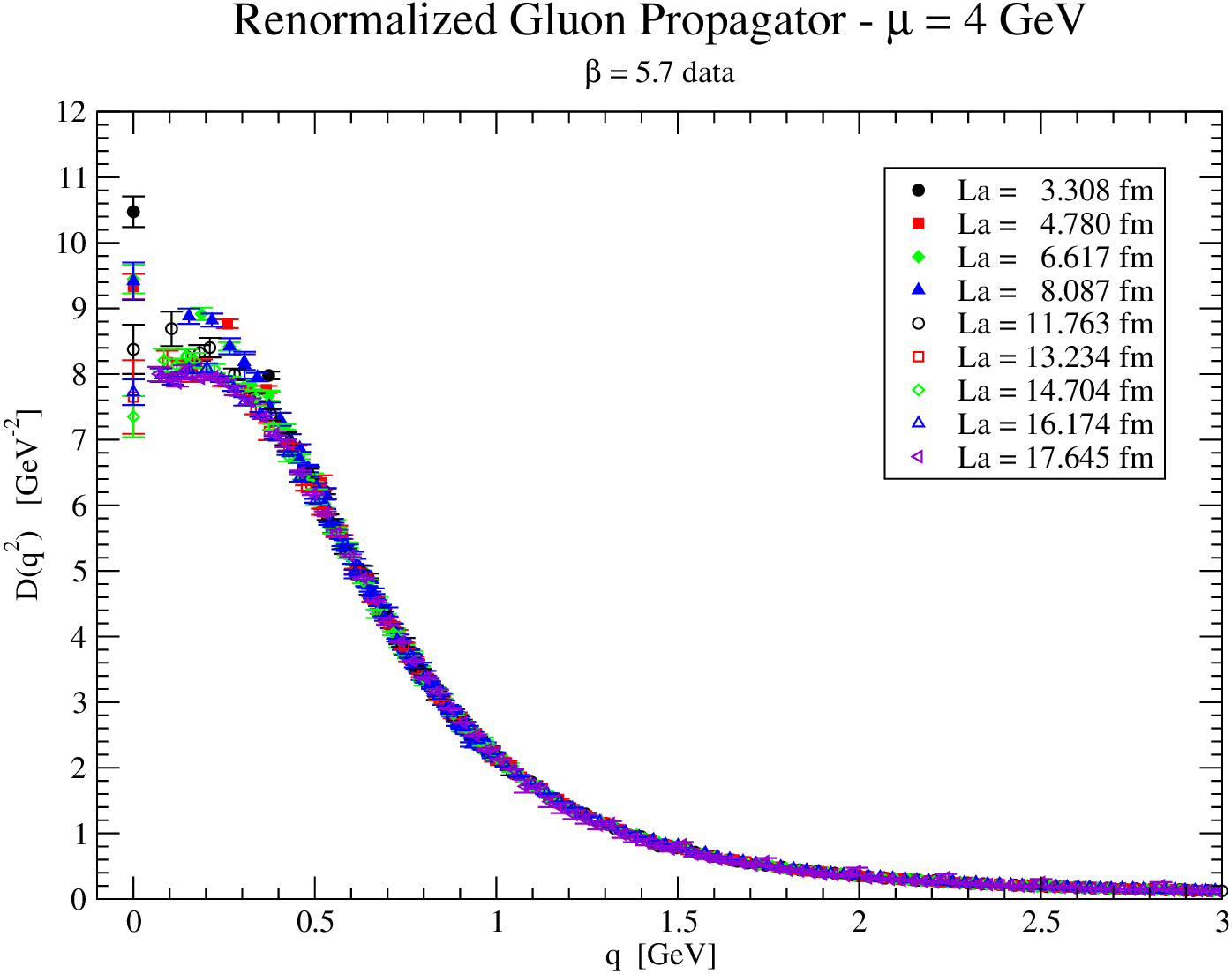} } \qquad
   \subfigure{ \includegraphics[scale=0.35]{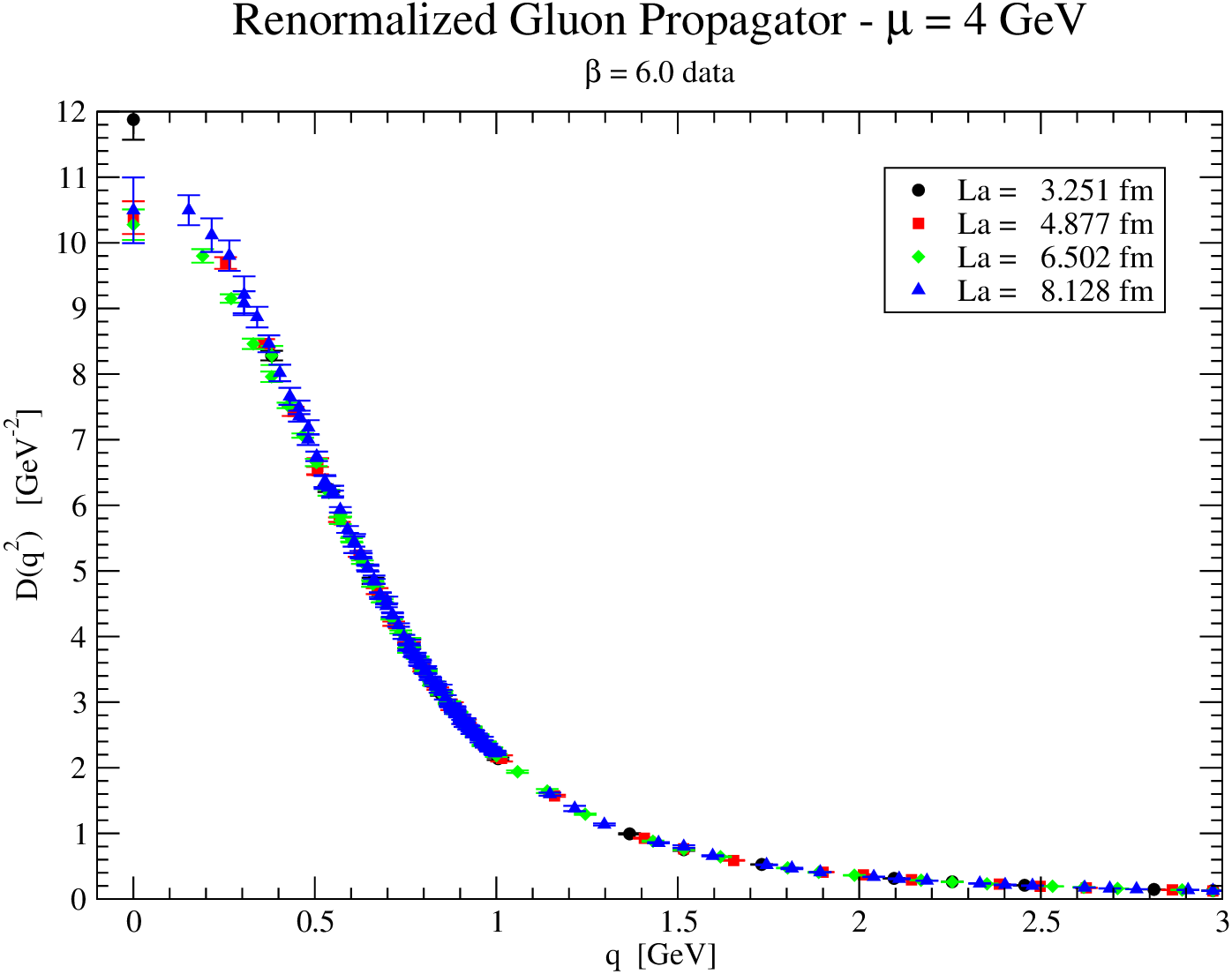} }

   \subfigure{ \includegraphics[scale=0.35]{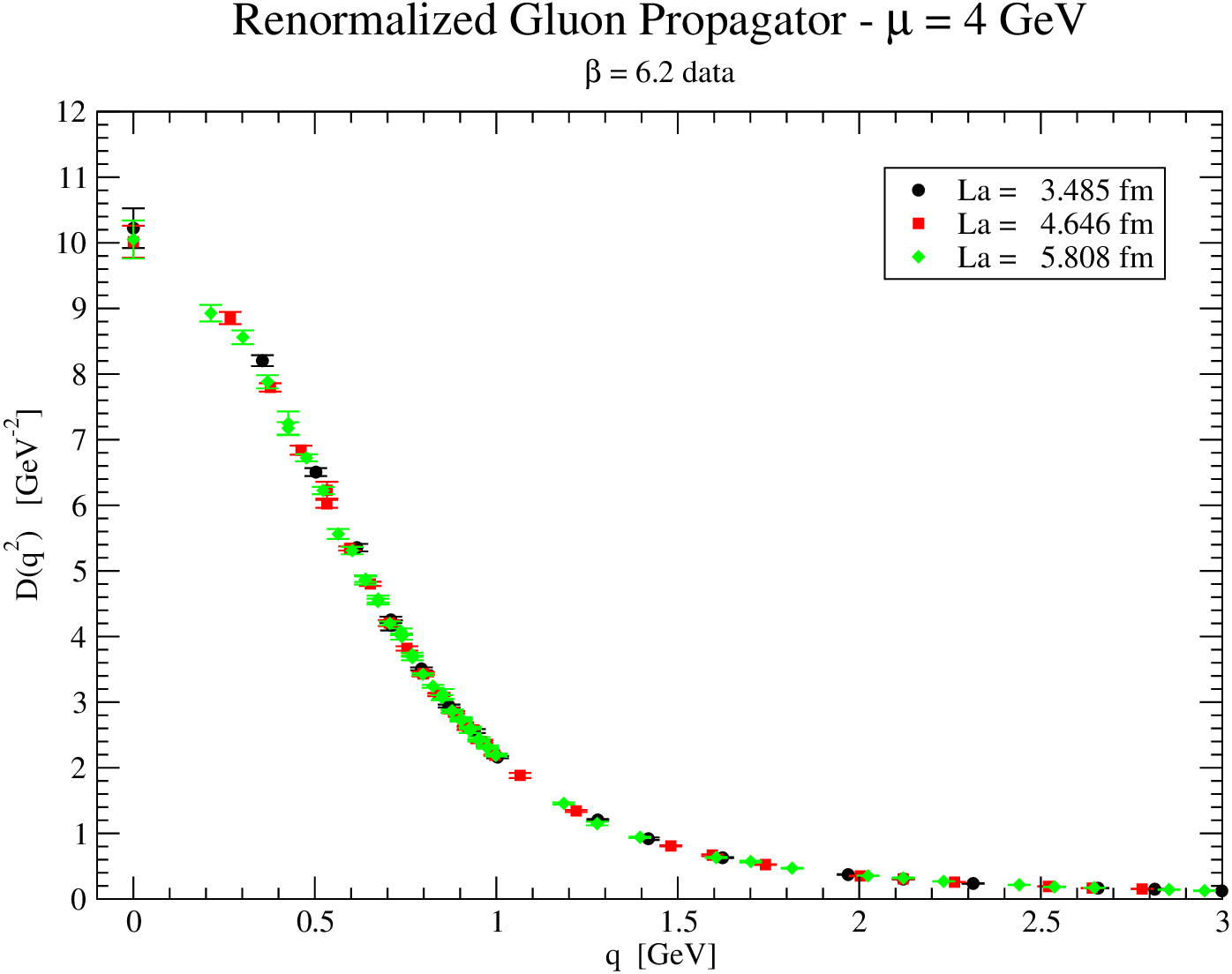} } \qquad
   \subfigure{ \includegraphics[scale=0.35]{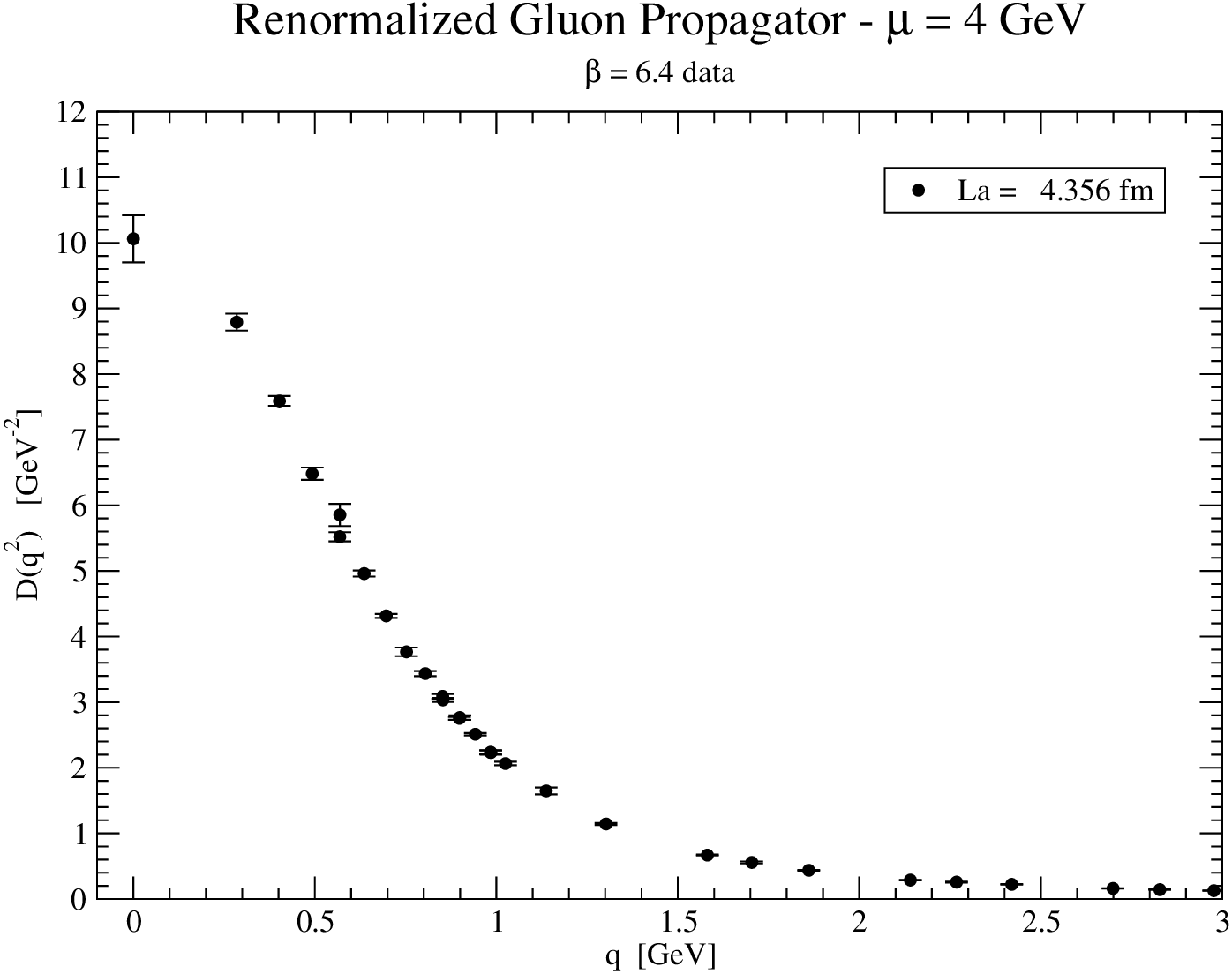} }
  \caption{Renormalized gluon propagator for $\mu = 4$ GeV for all lattice simulations.}
   \label{fig:glue4GeV}
\end{figure*}

\begin{figure*}[h] 
   \centering
   \subfigure{ \includegraphics[scale=0.35]{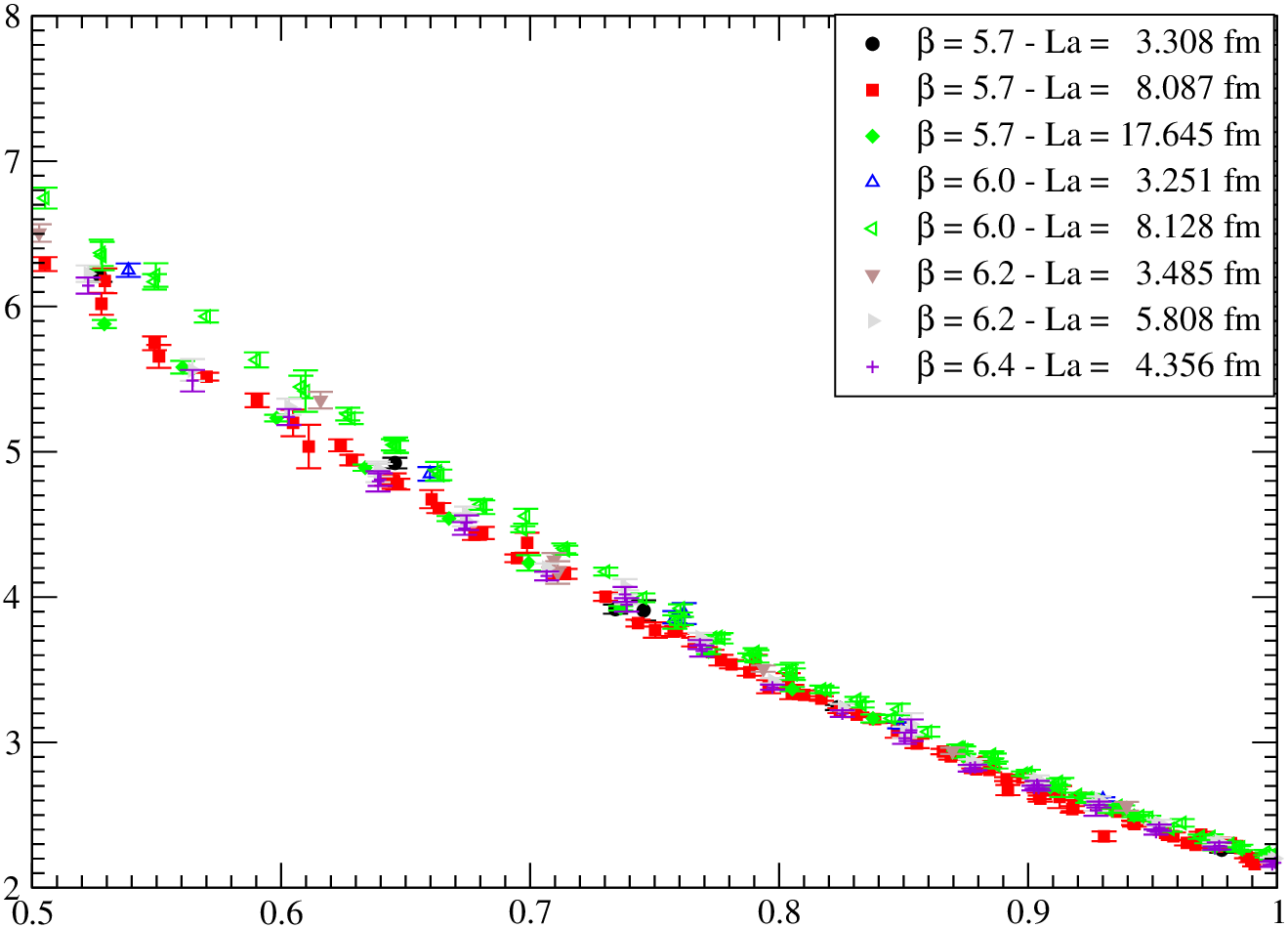} } \qquad
   \subfigure{ \includegraphics[scale=0.35]{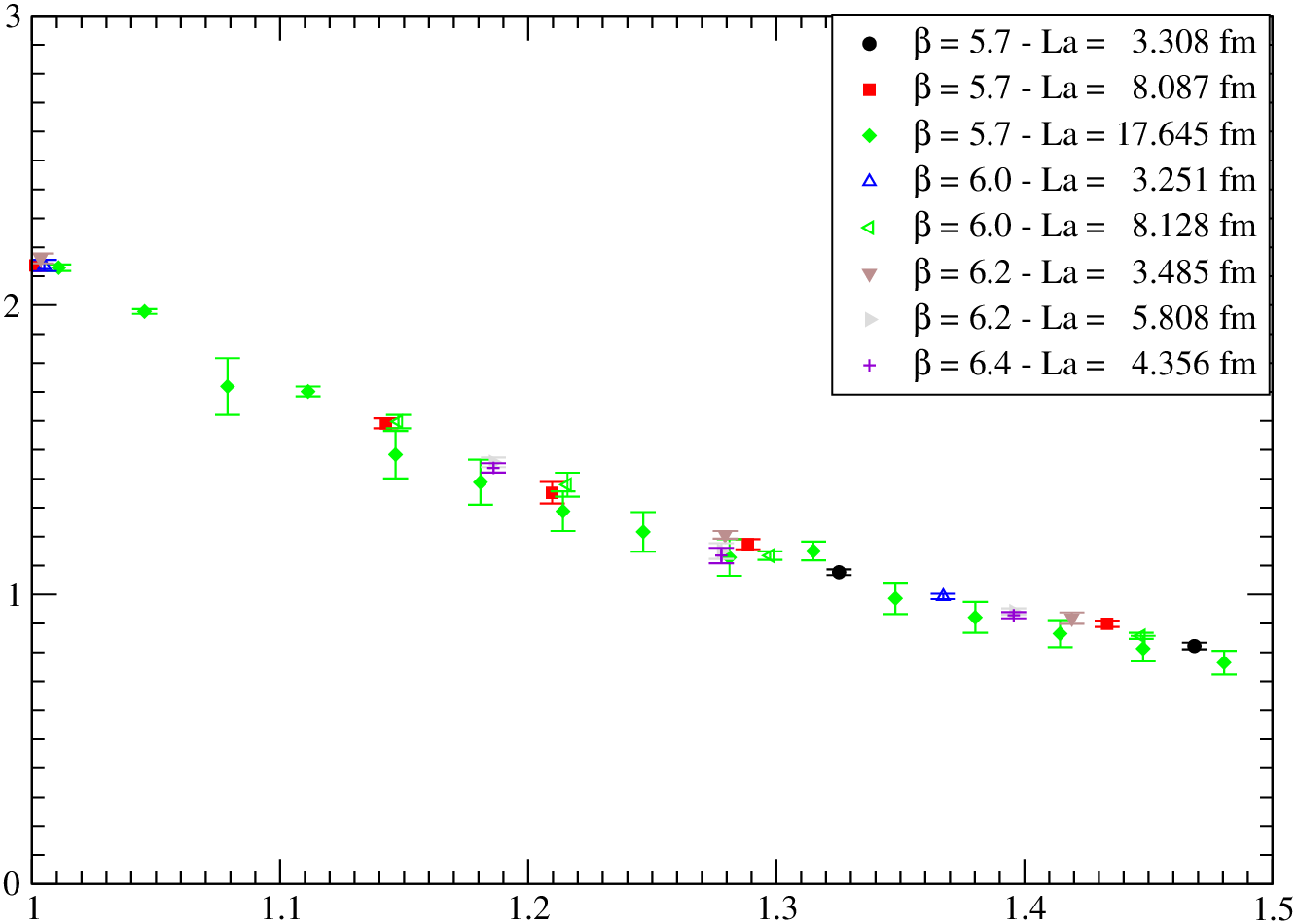} }

\vspace{0.65cm}
   \subfigure{ \includegraphics[scale=0.35]{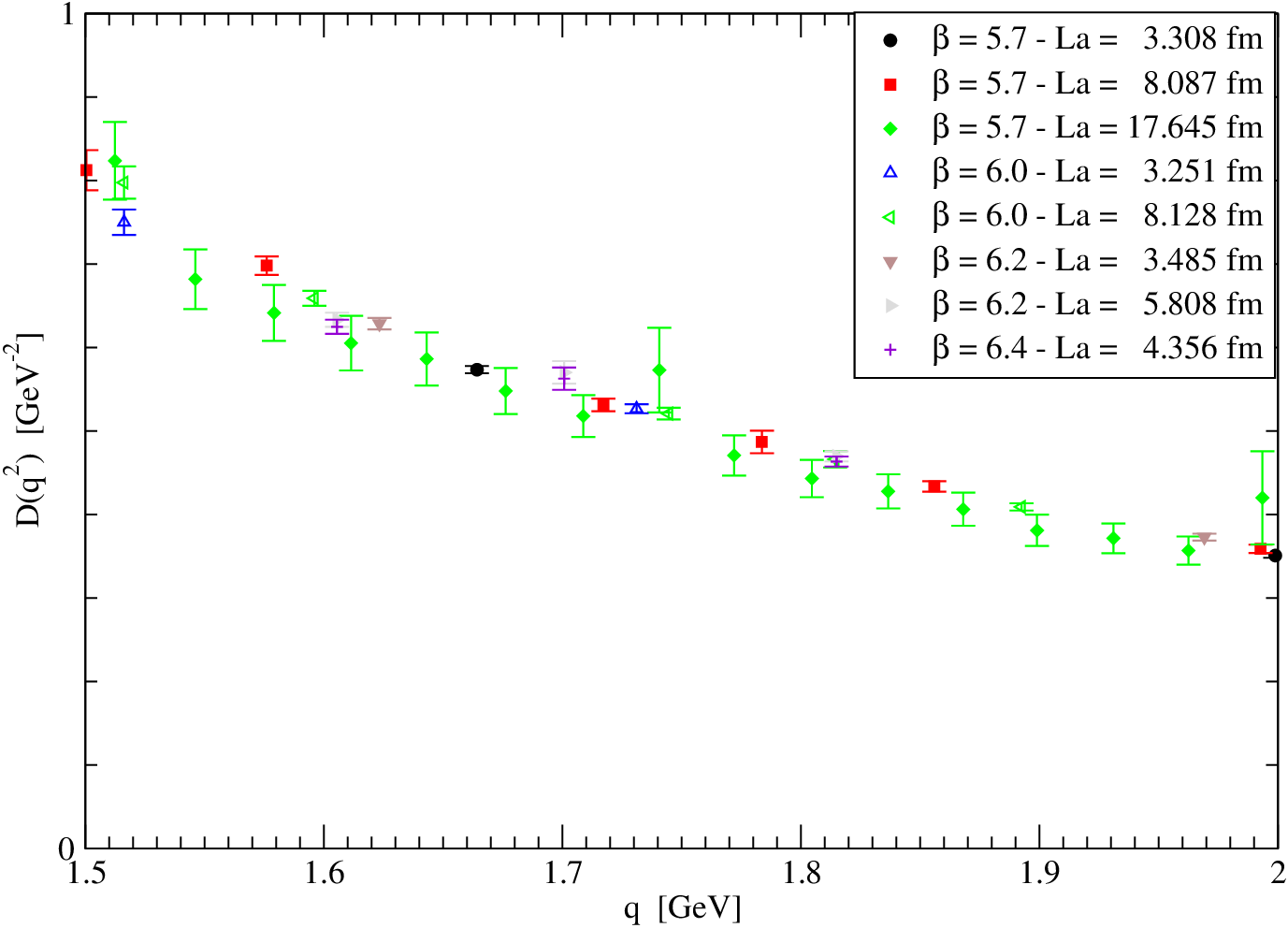} } \qquad
   \subfigure{ \includegraphics[scale=0.35]{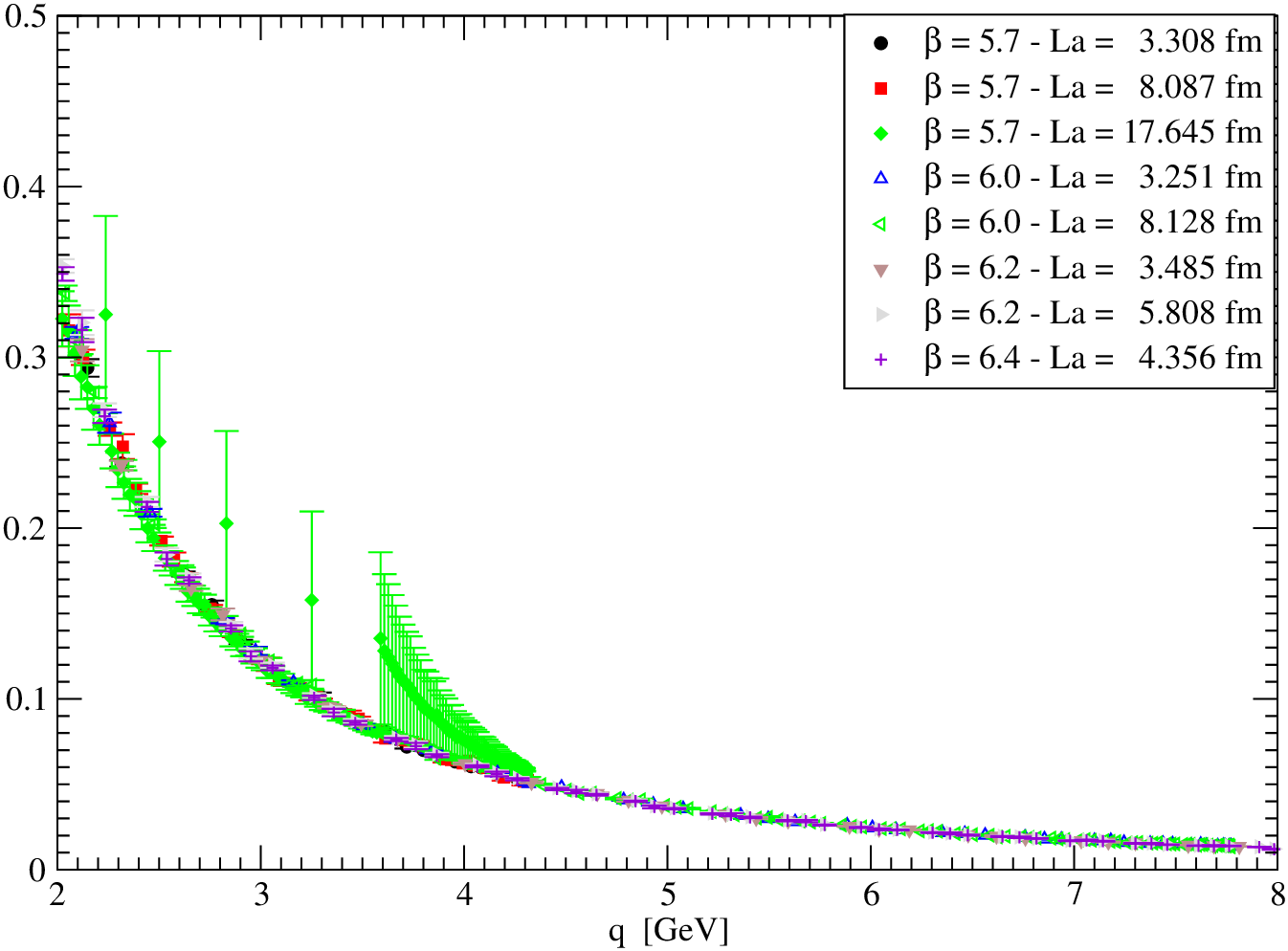} }
  \caption{Zoom of the renormalized gluon propagator for $\mu = 4$ GeV.}
   \label{fig:glue_zoom}
\end{figure*}

Despite the volume independence of $Z_R / a^2$, in the following,  to compute
the renormalized gluon propagator, instead of taking the central value
of (\ref{Eq:fitted_ZR}), we will use the central value obtained from fitting directly the lattice data to
(\ref{uvfit}). In this way, we hope to take into account possible systematics from using a limited statistics in each 
simulation. The renormalized gluon propagator for all ensembles can be seen in Fig. \ref{fig:glue4GeV}.
In Fig. \ref{fig:glue_zoom} we show the renormalized propagators, for a selected set of data, 
with a higher resolution in momentum scale. 

Figure \ref{fig:glue4GeV} shows that for a given lattice spacing the infrared gluon propagator 
is reduced as one goes towards larger physical volumes. This is well known and has been reported
many times. From the data at $\beta = 5.7$, the value of $D(0)$ decreases by a factor of $\sim 1.35$ 
when going from the smallest to the largest volume.

For larger momenta the lattice propagator is less dependent on the lattice volume and, as
Fig. \ref{fig:glue_zoom}  shows, for momenta above $\sim 900$ MeV the lattice gluon propagator
seems to be independent of the lattice volume, in the sense that all data sets define a unique curve for $D(q^2)$.
In this sense, one can claim that above such momenta the lattice propagator is free of lattice artifacts.


In what concerns the infrared region, no clear sign of a turnover on $D(q^2)$ is observed when approaching
$q = 0$. Therefore, one can claim that the lattice data points towards a finite and non-vanishing value for the 
zero momentum gluon propagator.

\section{On the Infinite Volume Limit of $D(0)$ \label{Volume}}

Let us now discuss the infinite volume limit for $D(0)$. The various simulations performed using
different $\beta$ values allow for independent extrapolations. 
For the extrapolation to the infinite volume we consider either
\begin{equation}
D(0) = \frac{c}{V} + D_\infty (0) \qquad\mbox{ or } \qquad D(0) = \frac{c}{L} + D_\infty (0) \, .
\label{Eq_D0_Ext}
\end{equation}
The fits of the lattice data for the different sets give
\begin{widetext}
\begin{displaymath}
 \begin{array}{l@{\hspace{0.9cm}}l@{\hspace{0.9cm}}l@{\hspace{1.5cm}}l@{\hspace{0.9cm}}l}
 \beta                & D_\infty (0) \mbox{ GeV}^{-2} & \chi^2/d.o.f. &D_\infty (0) \mbox{ GeV}^{-2} & \chi^2/d.o.f. \\
  5.7    & ~ ~ \, 9.26 ~ ~ ~ \, \pm ~0.13   & 0.8 & ~~~8.43 ~~ \pm ~ 0.61 & 2.6 \\
  5.7 \mbox{ \cite{largevolume} }  & ~ ~ \, 7.35 ~ \, ~ ~\pm ~0.30   & 1.8 & ~~~6.1 ~~~\, \,\pm ~1.4 & 1.3 \\
  6.0    & 10.157  ~ ~ \pm ~0.097            & 0.3 & ~~~8.79 ~~\pm ~0.64 & 1.7 \\
  6.2   & ~ ~ 9.980 ~~ \pm ~0.055               & 0.0 & ~~~9.72 ~~\pm ~ 0.25 & 0.1 
 \end{array}
\end{displaymath}
\end{widetext}
where the first columns of values refers to the extrapolation assuming a $1/V$ dependence and
the second columns assumes a $1/L$ linear function. 
The fits using combined data sets, i.e. all data or all $\beta = 5.7$ data,
give a too large $\chi^2/d.o.f.$, meaning that they are not described by any of the expressions in (\ref{Eq_D0_Ext}). 

\begin{figure}[t] 
   \centering
   \includegraphics[scale=0.35]{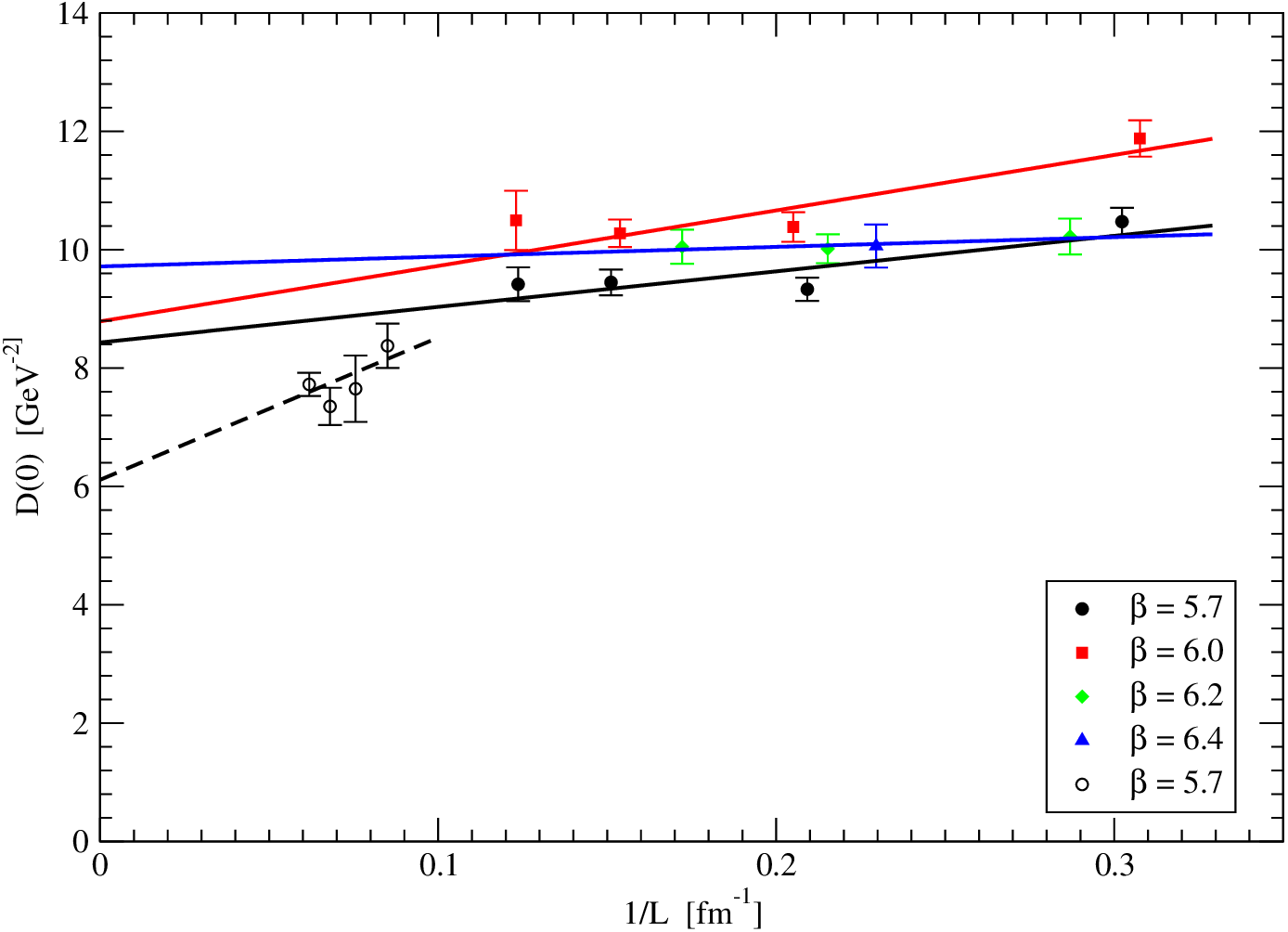} 
   \caption{Linear extrapolation of $D(0)$ to the infinite volume.}
   \label{fig:D0_infty}
\end{figure}

The fits and the linear extrapolation assuming a $1/L$ behavior can be seen in Fig.
\ref{fig:D0_infty}. Although the fits have acceptable $\chi^2/d.o.f$, as seen in Fig. \ref{fig:D0_infty}, the
linear extrapolation in $1/L$ does not provide a coherent picture of all the data sets. 
The situation does not improve when we consider a linear dependence in $1/V$. Note that the
linear extrapolation in $1/V$ gives larger values for $D(0)$, when compared to the $1/L$ extrapolation.

From the above results, one can claim a $D(0)$ in the range 6 -- 10 GeV$^{-2}$, in good agreement with the 
linear extrapolation performed by \cite{Bonnet01}, which used improved actions and gauge fixing,  
and where it was estimated a $D_\infty (0) = 7.95(13)$ GeV$^{-2}$. Recall that the extrapolation
performed in \cite{Bonnet01} assumed a $1/V$ linear dependence for the lattice data.

The analysis discussed so far supports a finite and non-vanishing $D(0)$. Indeed, no turnover of $D(q^2)$
is observed when approaching the zero momentum limit. However, as des\-cri\-bed in \cite{bounds} the scaling analysis
of the Cucchieri-Mendes bounds \cite{cucc3} and the ratios defined in \cite{ratios} do not exclude completely the possibility of 
having a vanishing zero momentum propagator.

\section{Finite Lattice Spacing Effects \label{spacing}}

The renormalization of the gluon propagator is devised to remove lattice spacing effects.
As can be observed in Fig. \ref{fig:glue_zoom}, the various renormalized lattice gluon propagators agree,
within errors, in the ultraviolet region but not necessarily at low momenta. 

The finite lattice spacing effects can be investigated comparing the renormalized gluon propagator 
computed using the same physical volume but different $\beta$ values. For the simulations reported
in Tab. \ref{tab:latsetup}, the following four sets have close physical volumes: 
\begin{itemize}
\item[(i)]
$(\beta = 5.7 , L = 3.308 \mbox{ fm}),   \\
  (\beta = 6.0 , L = 3.251 \mbox{ fm}), \\
  (\beta = 6.2 , L = 3.485 \mbox{ fm}) $;
\item[(ii)]
$(\beta = 5.7 , L = 4.780 \mbox{ fm}), \\
  (\beta = 6.0 , L = 4.877 \mbox{ fm}), \\
  (\beta = 6.2 , L = 4.646 \mbox{ fm}), \\
  (\beta = 6.4 , L = 4.359 \mbox{ fm});$
\item[(iii)]
$ (\beta = 5.7 , L = 6.617 \mbox{ fm}), \\
   (\beta = 6.0 , L = 6.502 \mbox{ fm});$
\item[(iiv)]
$ (\beta = 5.7 , L = 8.089 \mbox{ fm}), \\
   (\beta = 6.0 , L = 8.128 \mbox{ fm}).$ 
\end{itemize}
The physical volumes for the  simulations considered in each data set do not match perfectly. However,
the results summarized in Fig. \ref{fig:glue4GeV} show a very smooth dependence of the gluon
propagator with the lattice physical volume. Therefore, we expect that the conclusions drawn for
comparing the ensembles within each of the data sets illustrate how the lattice spacing changes $D(q^2)$.

\begin{figure*}[t] 
   \centering
   \subfigure{ \includegraphics[scale=0.35]{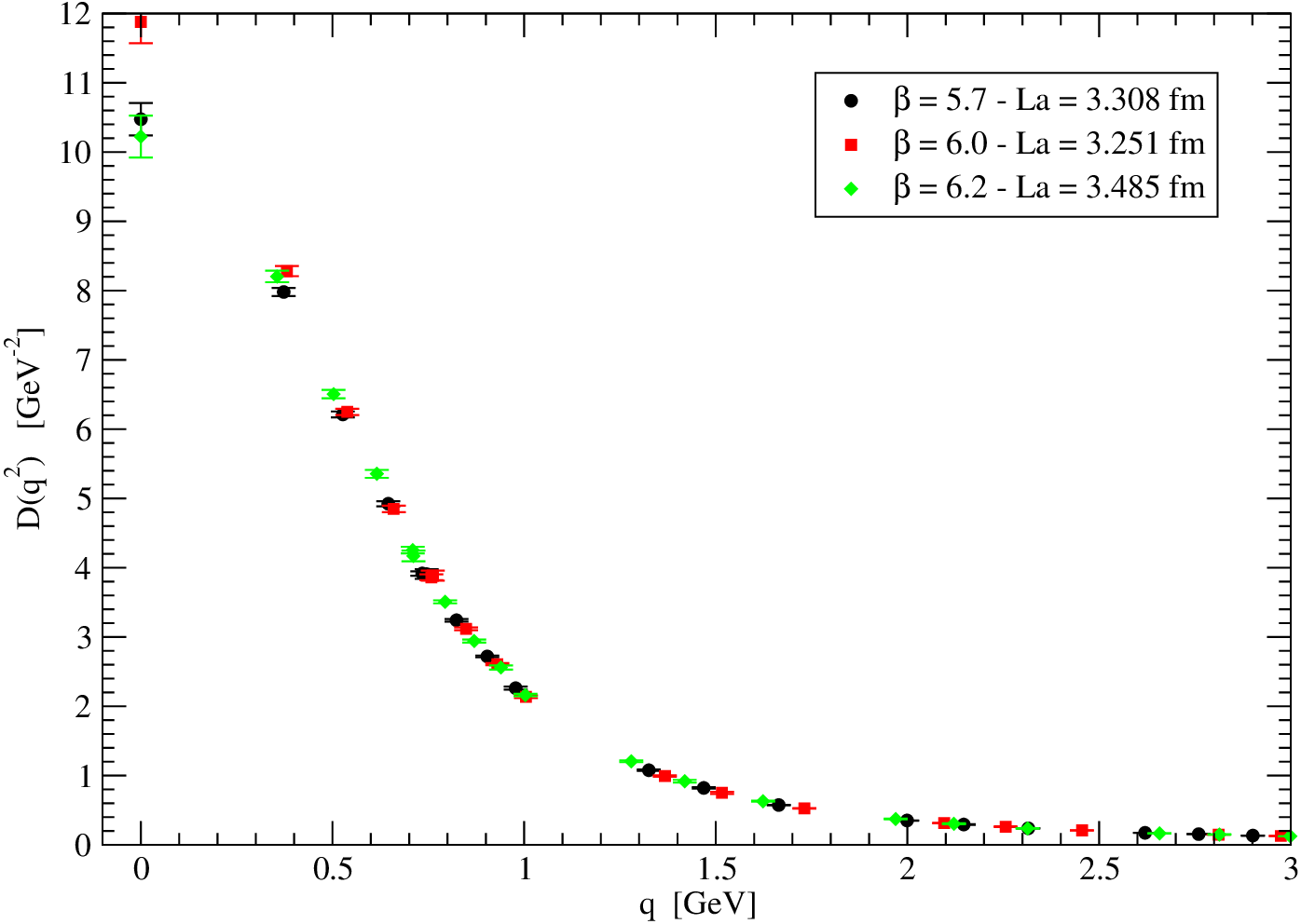} } \qquad
   \subfigure{ \includegraphics[scale=0.35]{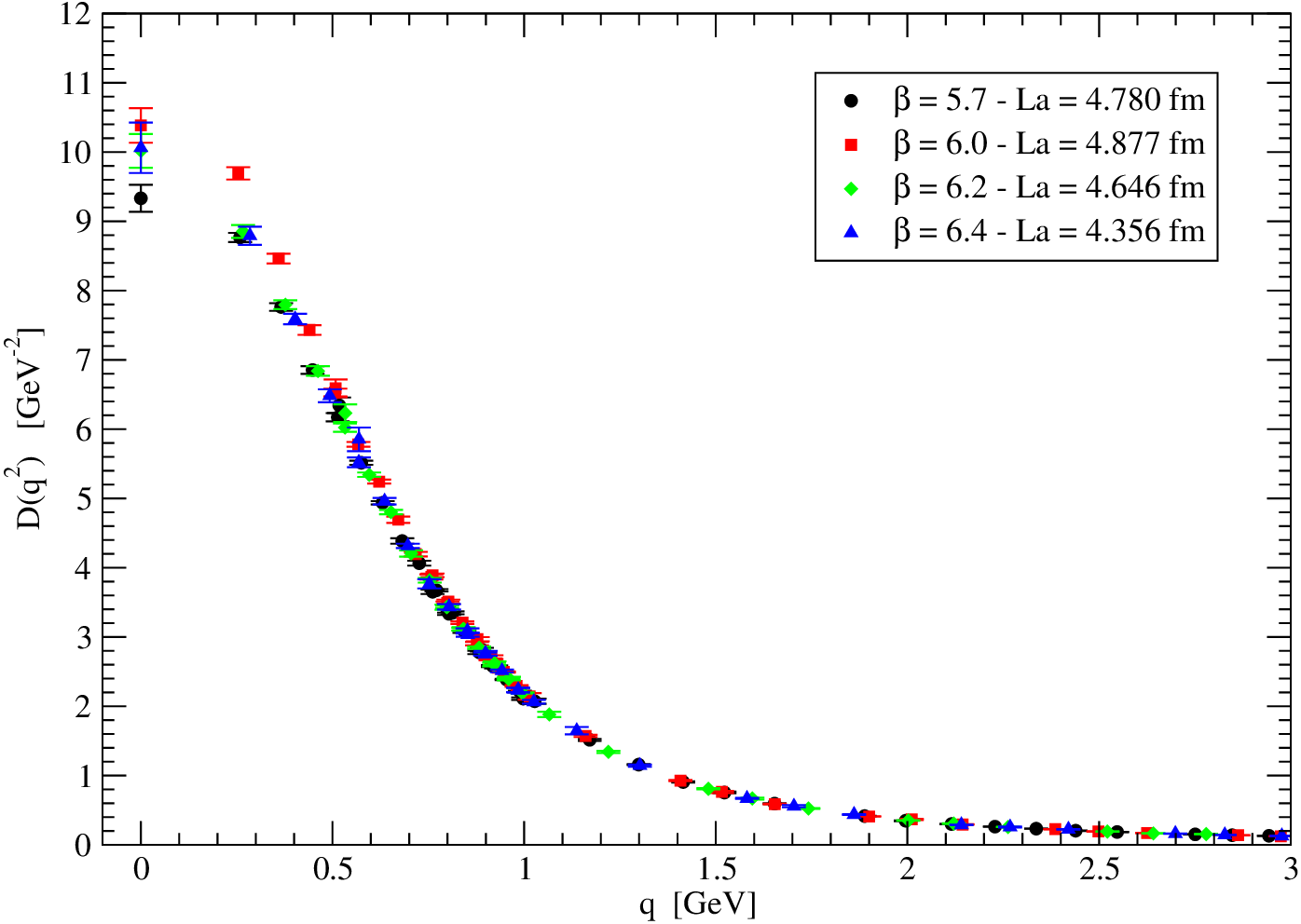} }

\vspace{0.6cm}
   \subfigure{ \includegraphics[scale=0.35]{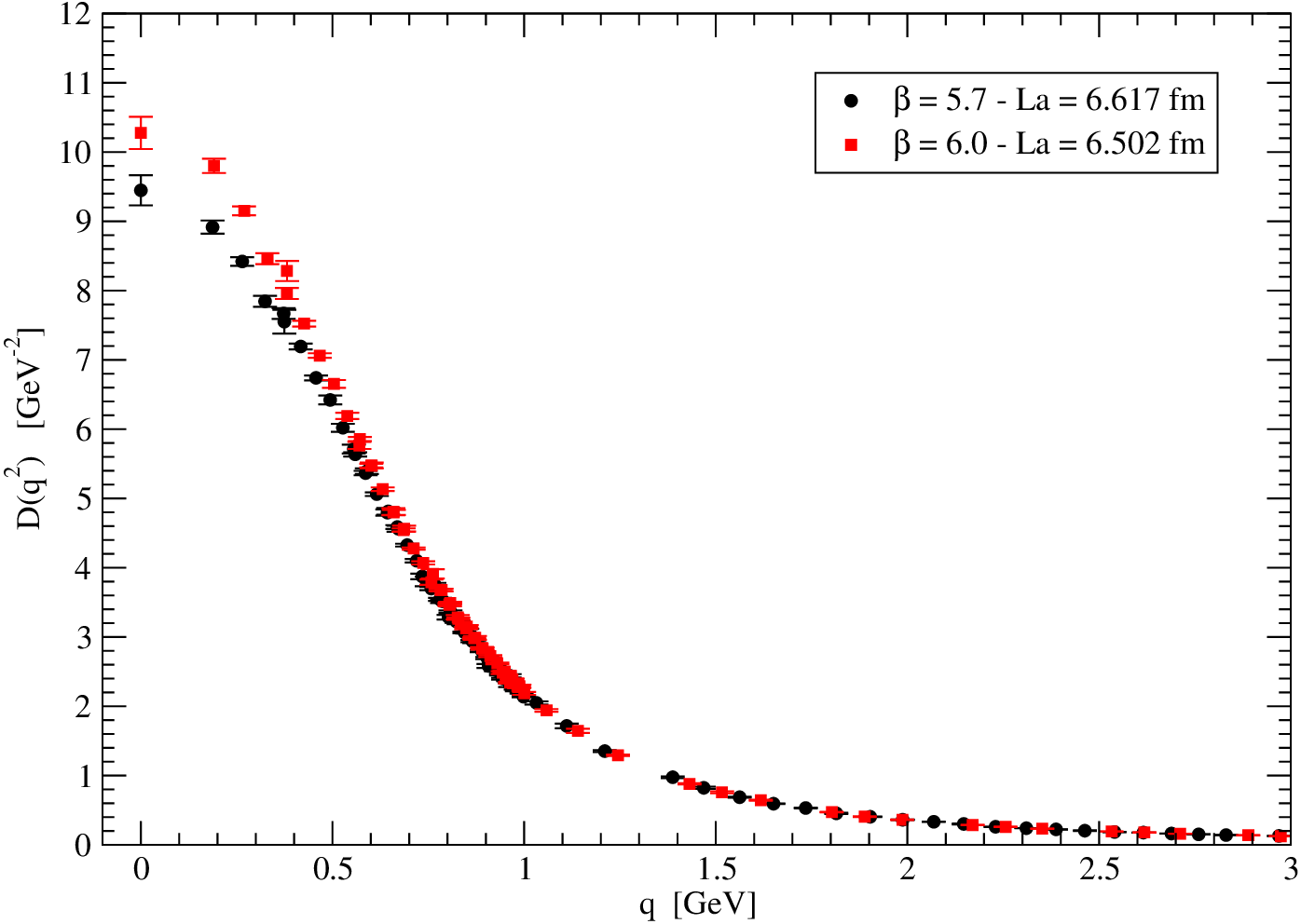} } \qquad
   \subfigure{ \includegraphics[scale=0.35]{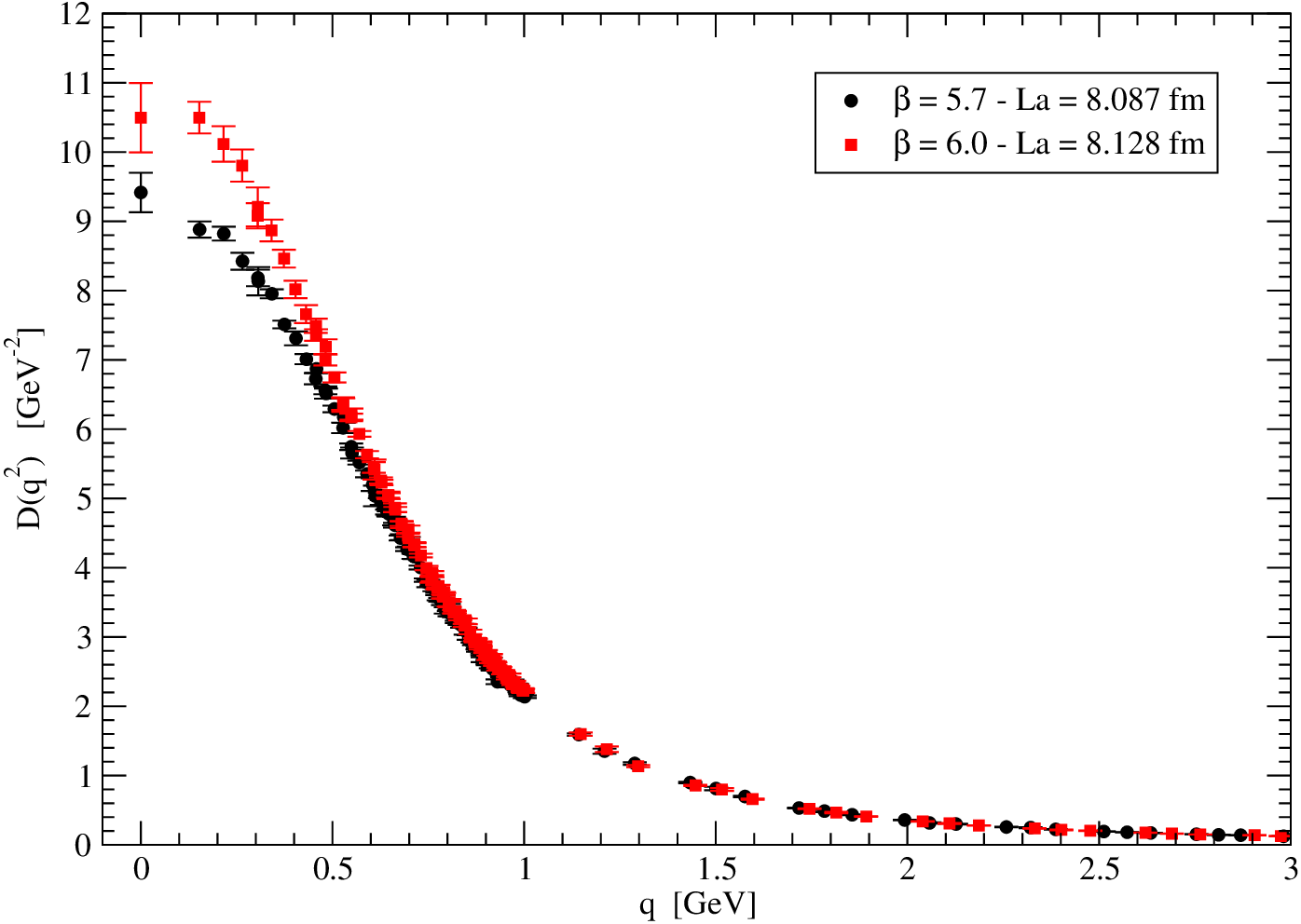} }
  \caption{Comparing the renormalized gluon propagator at $\mu = 4$ GeV for various lattice spacings
                and similar physical volumes.}
   \label{fig:glue_lattice}
\end{figure*}

In Fig. \ref{fig:glue_lattice} the data for the various sets are reported and compared.
From Fig.  \ref{fig:glue_lattice} one can see that for $q$ above $\sim 900$ MeV the lattice data define a unique
curve, i.e. the renormalization procedure removes all dependence on the ultraviolet cutoff $a$ 
for the mid and high momentum regions.

For the lower momenta Fig. \ref{fig:glue_lattice} shows that $D(q^2)$ is not independent of $a$. 
Further, comparing Figs. \ref{fig:glue4GeV} and \ref{fig:glue_lattice}  it follows that, in the infrared region, 
and for the lattice sizes $L = 3$ fm  or above, the corrections due to the finite lattice spacing seem to be 
larger than the corrections associated with the finite physical volume.  
Indeed, if, within each set of volumes, $D(0)$ seem to become compatible within two standard deviations,
the difference between propagators for the infrared region is clearly beyond two standard deviations. 
Furthermore,  Fig \ref{fig:glue_lattice} shows that the coarser lattice simulations,  i.e. those with smaller 
$\beta$ value, underestimate $D(q^2)$ in the infrared region. 
Note, however, that qualitatively the propagator is unchanged. 
In this sense, the large volume simulations performed by the 
Berlin-Moscow-Adelaide group provide a lower bound for the continuum infrared propagator.  

In appendix \ref{app}, we investigate how the choice of the renormalisation scale $\mu$ modifies 
the results reported in this section. Making different choices for $\mu$, we observe how $D(q^2)$ 
changes in various momenta regions.
For low momenta, it follows that the large lattice spacing simulation underestimates the propagator independently
of the renormalisation scale -- see the appendix for further details.

In what concerns the zero momentum gluon propagator, the previous analysis suggests that $D(0)$
computed at $\beta = 5.7$ gives a lower bound to its continuum value. 
For completeness, below we list 
the zero momentum gluon propagator for the largest physical volume associated with each $\beta$ value 
simulation
\begin{displaymath}
\begin{array}{l@{\hspace{0.7cm}}r@{\hspace{0.7cm}}r}
  \beta & La ~(\mbox{fm}) & D(0) ~(\mbox{GeV}^{-2})\\
  5.7    & 8.087    &  9.42 \pm 0.28 \\
  5.7    & 16.178  &  7.72 \pm  0.20 \\
  6.0    & 8.128    & 10.50 \pm  0.50 \\
  6.2    & 5.808    & 10.05 \pm  0.29 \\
  6.4    & 4.359    & 10.06 \pm  0.36
\end{array}  
\end{displaymath}

\section{Modelling the Gluon Propagator and Removing the Lattice Artifacts \label{glue_D0}}

As described in the previous sections, the renormalized gluon propagator depends on the physical 
lattice volume, for the same lattice spacing, and on the lattice spacing, for the same physical volume. 
In what concerns the infrared region, it is observed that the propagator decreases slowly when the lattice 
physical volume is increased, while keeping the same lattice spacing. On the other hand, when the physical 
volume is keep constant, the infrared propagator is enhanced when the lattice spacing decreases. 
According to the results described above, the dominant effect is associated with the lattice spacing 
rather than with the volume dependence.

In the present work we aim to provide a gluon propagator which is free of finite size effects.
Such a goal can be achieved, at least partially, through modeling the propagator. Of the possible 
functional forms, we chose to fit the lattice data to 
\begin{equation}
  D(q^2) = Z \frac{q^2 + M^2_1}{q^4 + M^2_2 \, q^2 + M^4_3} \, ,
  \label{Eq:fit_function}
\end{equation}  
where $Z$ is dimensionless and $M_1$, $M_2$ and $M_3$ have dimensions of mass. 
The above expression is the tree level prediction of the so-called refined Gribov-Zwanziger action supplied by
an extra fitting normalization parameter $Z$. For $Z = 1$, as shown in \cite{Dudal2010}, Eq. (\ref{Eq:fit_function})
describes the lattice gluon propagator in the deep infrared and up to momenta $\sim 1.5$ GeV.
Although in the framework of the refined Gribov-Zwanziger action
the different mass parameters have a precise meaning and are associated with various condensates and the 
Gribov mass parameter, here we do not explore such a connection. 
Besides the theoretical motivation coming from the refined Gribov-Zwanziger action, (\ref{Eq:fit_function})  
can be viewed as a Pad\' e approximation to the lattice data. 

\begin{figure}[t] 
   \centering
   \includegraphics[scale=0.35]{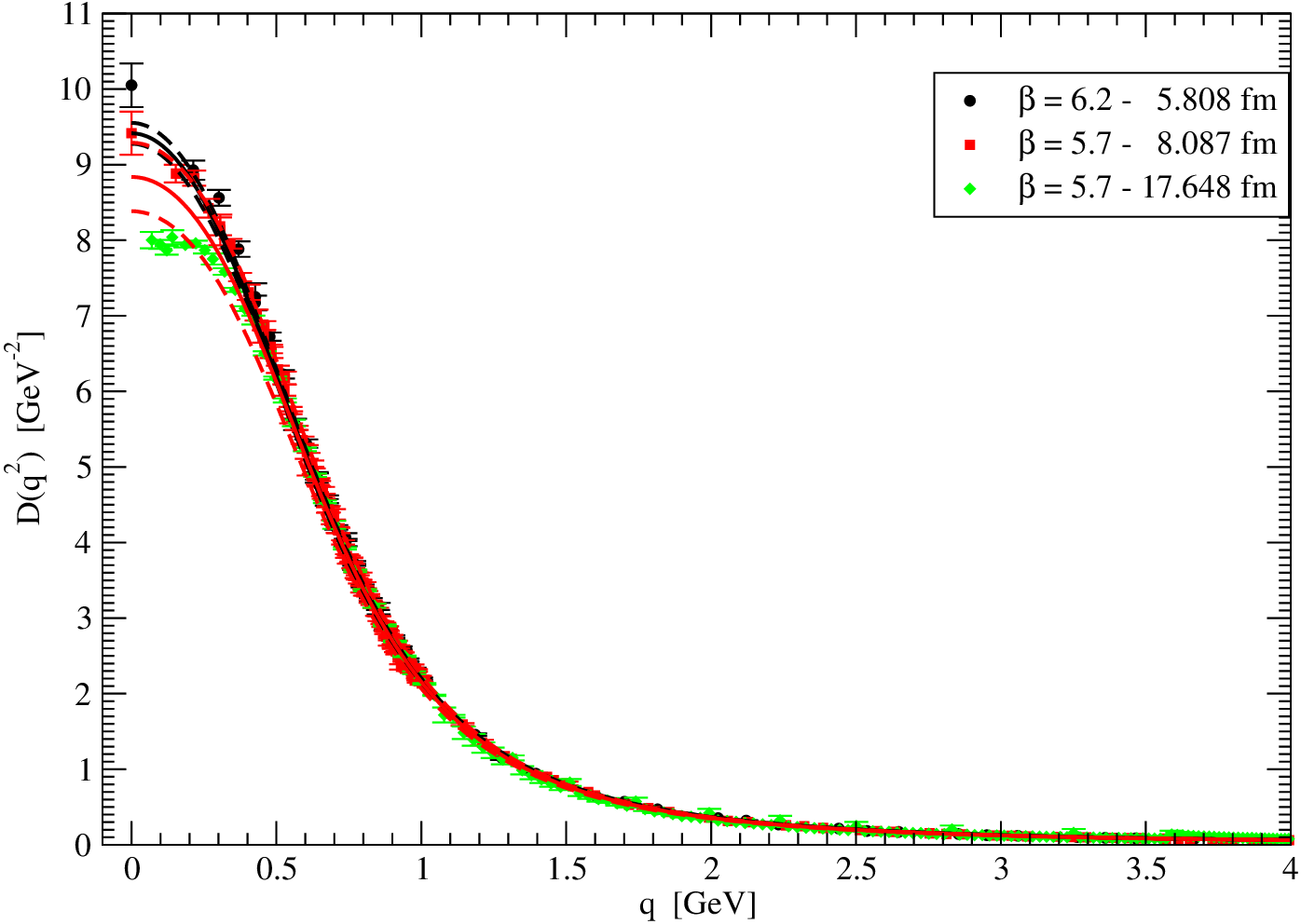} 
   \caption{Extrapolation of the gluon propagator to the infinite volume limit for the $\beta = 5.7$ and $\beta = 6.2$
                  ensembles. The full line represents the $D(q^2)$ computed with (\ref{Eq:fit_function}) and using
                  the central values of the extrapolated parameters. The dashed lines are the one standard deviations
                  for $D(q^2)$ computed assuming gaussian error propagation.}
   \label{fig:glue_ext}
\end{figure}

The functional form (\ref{Eq:fit_function}) does not include any log terms generated by the loop
corrections. Therefore, we do not expect that it describes the full set of momenta. Being so, we have to
define a fitting range, which includes the infrared region and goes beyond the 1 GeV, where all data sets
define a unique curve, and where (\ref{Eq:fit_function}) is able to reproduce properly the lattice propagator.
Although in our analysis we have considered various fittings ranges, in the following we will report 
only the results for a fitting range starting at $q = 0$ and going up to $q_{max} = 4$ GeV. 
The results for the other fitting ranges being similar. The outcome of the fits being
\begin{widetext}
\begin{displaymath}
  \begin{array}{ll@{\hspace{0.7cm}}l@{\hspace{0.7cm}}l@{\hspace{0.7cm}}l@{\hspace{0.7cm}}l@{\hspace{0.7cm}}l}
  \beta  &  L  &  Z  & M^2_1  & M^2_2  & M^4_3  & \chi^2/d.o.f \\
  5.7     & 18 &  0.8290(91) & 3.94(18) & 0.583(41) & 0.3224(89) & 1.86 \\
            & 26 & 0.8156(74)  & 4.16(15) & 0.603(32) & 0.3533(78) & 2.32 \\
            & 36 & 0.8254(62)  & 3.98(11)  & 0.557(23) & 0.3523(58) &1.96 \\
            & 44 & 0.8231(71)  & 4.08(12)  & 0.583(25) & 0.3569(64) & 1.65 \\
   6.0    & 32 & 0.811(11)     & 4.45(22)  & 0.694(48) & 0.3243(94) & 1.15 \\
            & 48 & 0.7996(99)  & 4.51(19)  & 0.669(37)  & 0.3384(83) & 1.81 \\
            & 64 & 0.8209(64)  & 4.21(11)  & 0.604(22)  & 0.3362(50) & 1.42 \\
            & 80 & 0.8196(59)  & 4.22(10)  & 0.631(22)  & 0.3177(41) & 0.81 \\
   6.2    & 48 & 0.806(14)    & 4.32(25)  & 0.623(49)  & 0.342(11)   &1.25 \\
            & 64 & 0.8135(75)  & 4.36(14)  & 0.653(28)  & 0.3575(68) &0.79 \\
            & 80 & 0.8169(96)  & 4.38(17)  & 0.654(34)  & 0.3635(82) &1.15
  \end{array}
\end{displaymath}
\end{widetext}
where all the mass parameters are given in powers of GeV. For each $\beta$ set we combine all volumes
and perform a linear extrapolation in $1/L$ to the infinite volume of each parameter independently. 
The $\chi^2/d.o.f$ for the $Z$, $M^2_1$, $M^2_2$, $M^4_3$ associated with the extrapolation being,
respectively, 0.78, 0.65, 0.64, 1.01 for the $\beta = 5.7$ ensembles,
1.23, 0.56, 1.05, 4.61 for the $\beta = 6.0$ ensembles and
0.00, 0.00, 0.07, 0.04 for the $\beta = 6.2$ ensembles. The extrapolated parameters being
\begin{displaymath}
  \begin{array}{l@{\hspace{0.7cm}}l@{\hspace{0.7cm}}l@{\hspace{0.7cm}}l}
  \beta =  &  5.7          &  6.0          & 6.2 \\
  Z           & 0.821(10) & 0.830(13) & 0.83333(17) \\
 M^2_1   & 4.09(17)   & 4.01(16)   & 4.473(21) \\
 M^2_2   & 0.558(36) & 0.565(46) & 0.704(29) \\
 M^4_3   & 0.380(11)  & --              & 0.3959(54) \\
 D(0)      & 8.84(45)     & --              & 9.42(14) \\
  \end{array}
\end{displaymath}
where $D(0)$ is given in GeV$^{-2}$. 
The extrapolated propagators together with the largest volume simulations results for each $\beta$ are shown
in Fig. \ref{fig:glue_ext}. Note that due to the bad extrapolation for $M^4_3$ when using the $\beta = 6.0$ 
ensembles, Fig. \ref{fig:glue_ext} does not include any information coming from these data.

From Fig. \ref{fig:glue_ext}, comparing the extrapolated $\beta = 5.7$ propagator using volumes up to 
(8.1 fm)$^4$ with the largest volume simulated by the Berlin-Moscow-Adelaide group, 
it follows that the two propagators are compatible within two standard deviations. The extrapolated propagator
being above the (17.1 fm)$^4$ lattice simulation results.
This supports, once more,
the previous observation that $\beta = 5.7$ results provides a lower bound for the infinite volume propagator.
Furthermore,
the extrapolated propagator from the $\beta = 6.2$ data is above the extrapolated $\beta = 5.7$ propagator. Note,
however, that the extrapolated $D(q^2)$ are compatible within two standard deviations.

In what concerns the infinite volume limit for $D(0)$, the extrapolations give a value of 8.84(45) GeV$^{-2}$
from the $\beta = 5.7$ data and 9.42(14) GeV$^{-2}$ from the $\beta = 6.2$ data. These values
are slightly above the infinite volume limit $D_\infty (0) = 7.95(13)$ GeV$^{-2}$
estimated in \cite{Bonnet01}.

\section{Conclusions \label{resultados}}

In this work, the effects of using finite volume and finite lattice spacing to compute the Landau gauge
gluon propagator in lattice QCD simulations are investigated. The propagator was computed for various
physical volumes, all above (3 fm)$^4$, and various lattice spacings ranging from 0.18 fm down to 0.054 fm.
Our simulations confirm, once more, that the infrared propagator decreases as the lattice volume increases.
Furthermore, the comparison between the data generated at similar physical volumes but different lattice
spacings, show that, in what concerns the infrared momenta, the finite lattice spacing effects are larger than 
the finite volume effects. Our analysis shows that decreasing the lattice spacing leads to an increase in 
$D(q^2)$ in the low momenta region. In this sense, the data coming from the large volume simulations by
the Berlin-Moscow-Adelaide group, which have an $a \approx 0.18$ fm, provides a lower bound for the continuum
propagator. In principle, the same behavior is expected for the large volume simulations performed for the
SU(2) gauge group where $a \approx 0.22$ fm. This observation is also supported by extrapolations of the 
lattice data to the infinite volume limit.

Besides the finite lattice and finite volume effects, we also investigate the zero momentum gluon propagator.
The various extrapolations performed here point towards a finite value for
 $\left. D(0) \right|_{\mu = 4 \mbox{GeV}}$ around 9 GeV$^{-2}$.

\section{Acknowledgments}

We would like to thank the Berlin-Moscow-Adelaide group for sending us their
data and for allowing to use it. Simulations have been carried out 
in Milipeia and Centaurus computer clusters at the University
of Coimbra. Paulo Silva is supported by FCT under contract SFRH/BPD/40998/2007.
Work supported by projects CERN/FP/123612/2011, CERN/FP/123620/2011 and 
PTDC/FIS/100968/2008, projects developed under the initiative QREN financed 
by the UE/FEDER through the Programme COMPETE - “Programa Operacional 
Factores de Competitividade”.

\begin{appendix}
\section{Renormalization and Finite Size Effects \label{app}}

In section \ref{spacing} the propagators renormalised at $\mu = 4$ GeV were compared for the same
physical volume and various lattice spacings. 
 As seen in Fig. \ref{fig:glue_lattice}, for the infrared gluon propagator, the effects of using a large lattice spacing are much larger than those associated with the physical volume of the box where the simulation is performed. 
What we want to address here, is how robust are the results of Fig. \ref{fig:glue_lattice}, 
relative to a change of the renormalisation scale $\mu$.

In order to be able to renormalise the propagator at low momenta, we fit the bare gluon propagator to
\begin{equation}
   D(q^2) = Z \frac{q^2 + M^2_1}{q^4 + M^2_2 q^2 + M^4_3} \nonumber 
\end{equation}
up to $q \sim 4$ GeV. For each value of $\mu$, the renormalised propagator is defined as in 
Eqs. (\ref{Eq:DRDLAT}) and (\ref{Eq:Rpoint}). In the following, we will consider two cases $\mu = 500$ MeV and
$\mu = 1$ GeV and study how the propagators differ. Furthermore, to avoid the inclusion of a large number of figures,
we will show only the results for the simulations performed at $La \approx 8$ fm.
The motivation for this choice being that this physical volume has the largest collection of momenta below $q = 1$ GeV.

\begin{figure*}[t] 
   \centering
   \subfigure{ \includegraphics[scale=0.35]{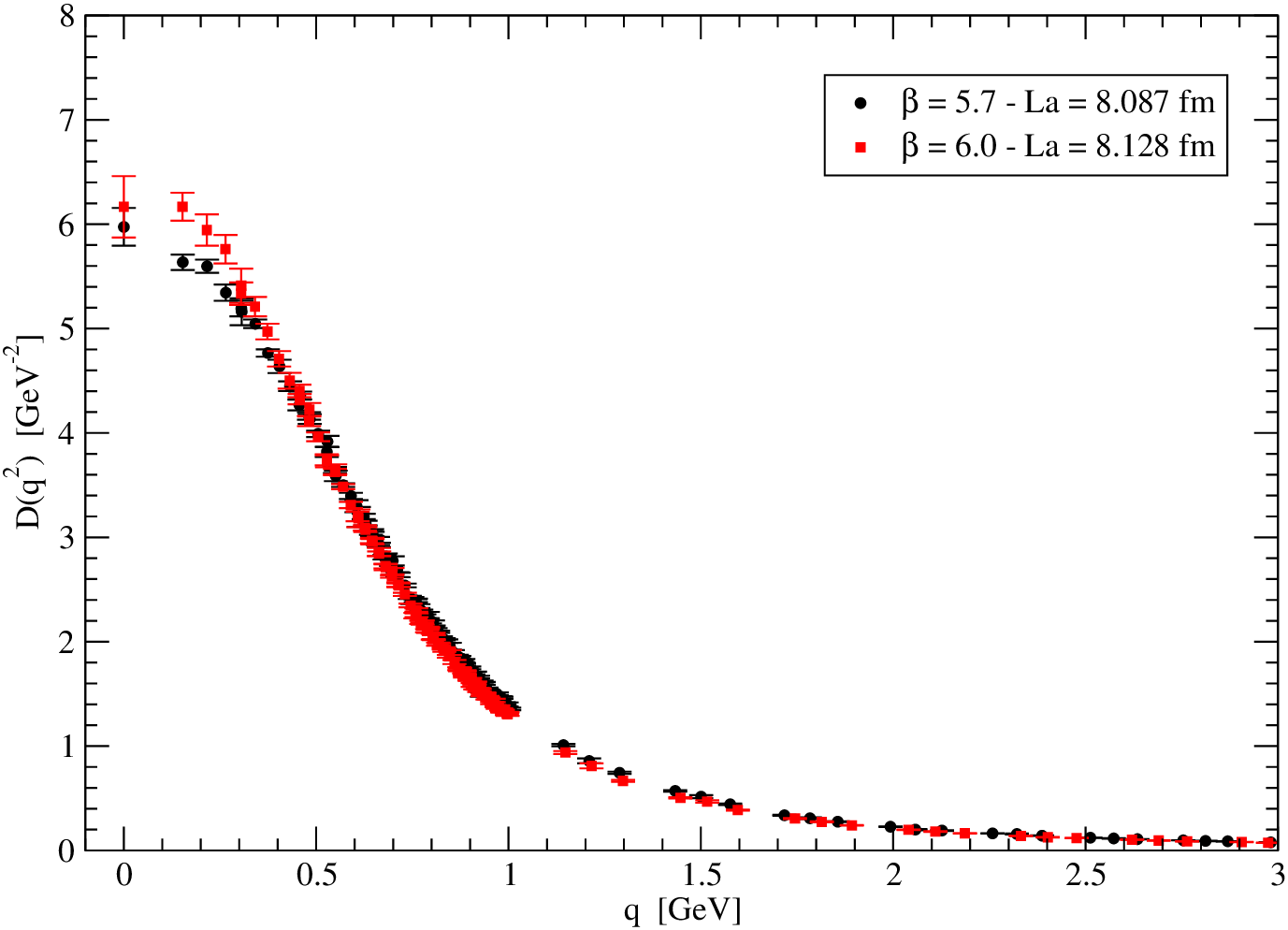} } \qquad
   \subfigure{ \includegraphics[scale=0.35]{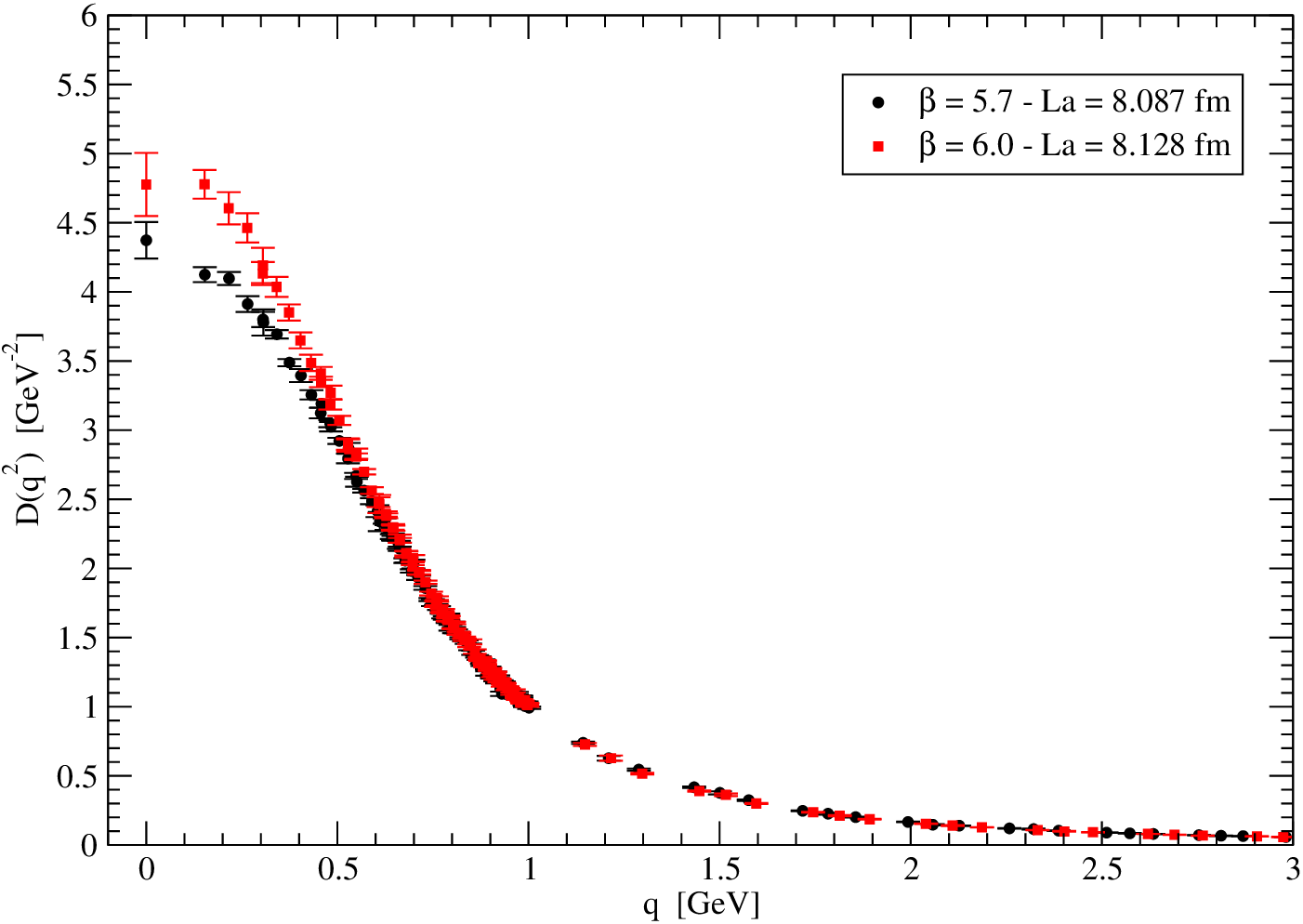} }
  \caption{Gluon propagator renormalised at $\mu = 500$ MeV (left) and at $\mu = 1$ GeV  (right).}
   \label{fig:glue1GeV500MeV}
\end{figure*}

The propagators can be seen in Fig. \ref{fig:glue1GeV500MeV} for momenta up to $q = 3$ GeV. 
The figure shows that, for both renormalisation points, the differences in the infrared region are well beyond 
one standard deviation and, in this sense, confirm the results reported in section \ref{spacing}.

\begin{figure*}[t] 
   \centering
   \subfigure{ \includegraphics[scale=0.35]{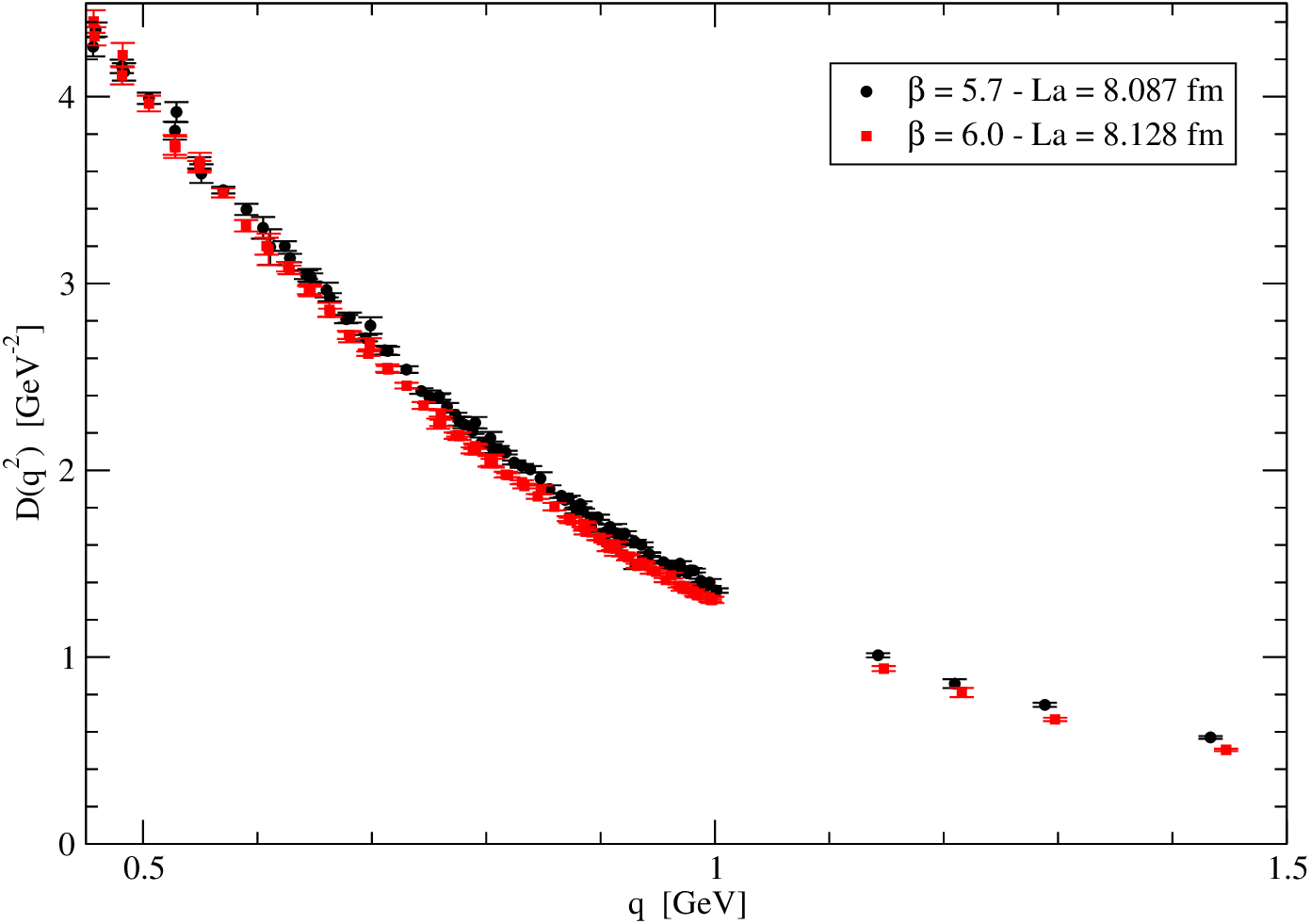} }\qquad
   \subfigure{ \includegraphics[scale=0.35]{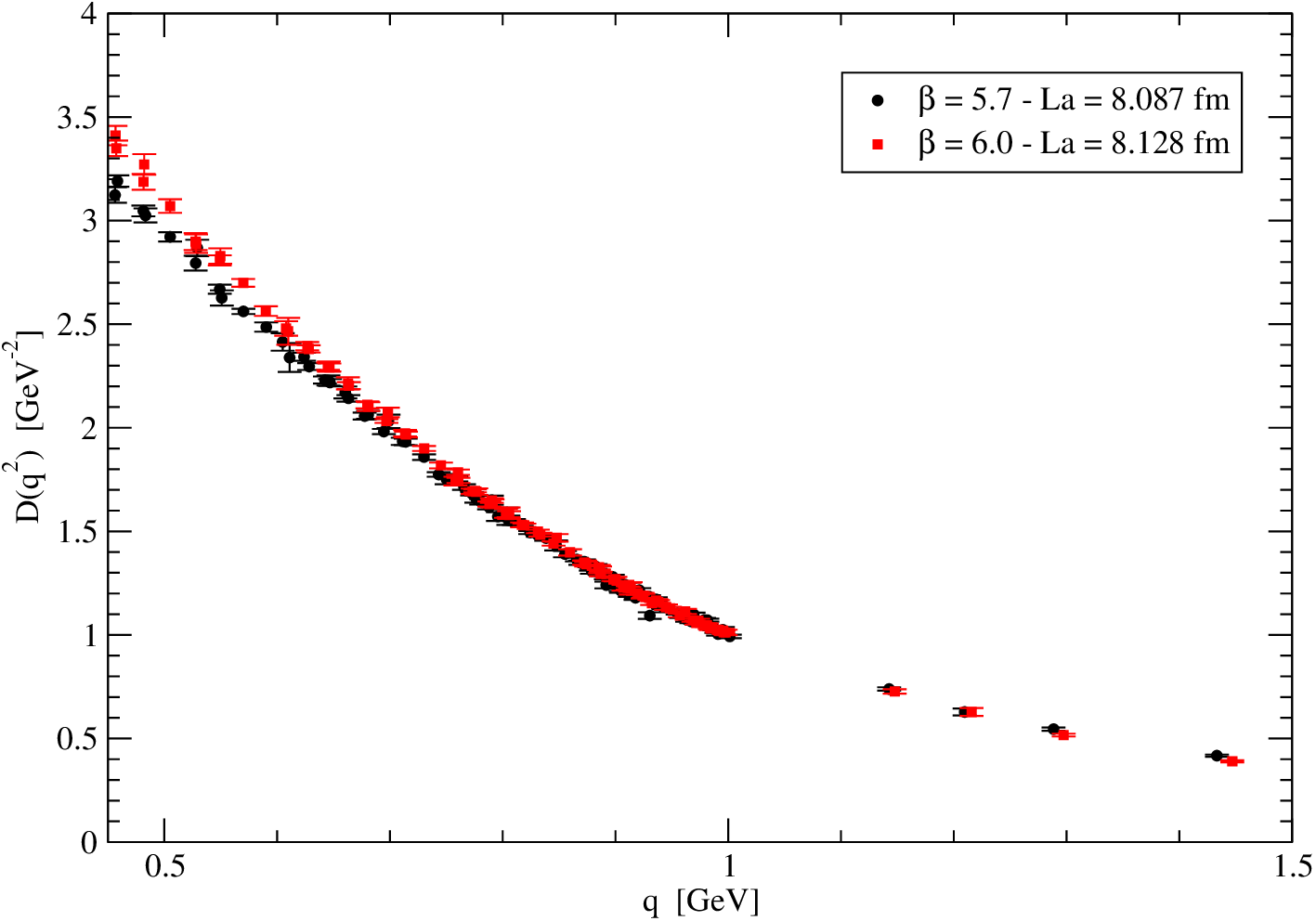}  }
   
   \vspace{0.5cm}
   \subfigure{ \includegraphics[scale=0.35]{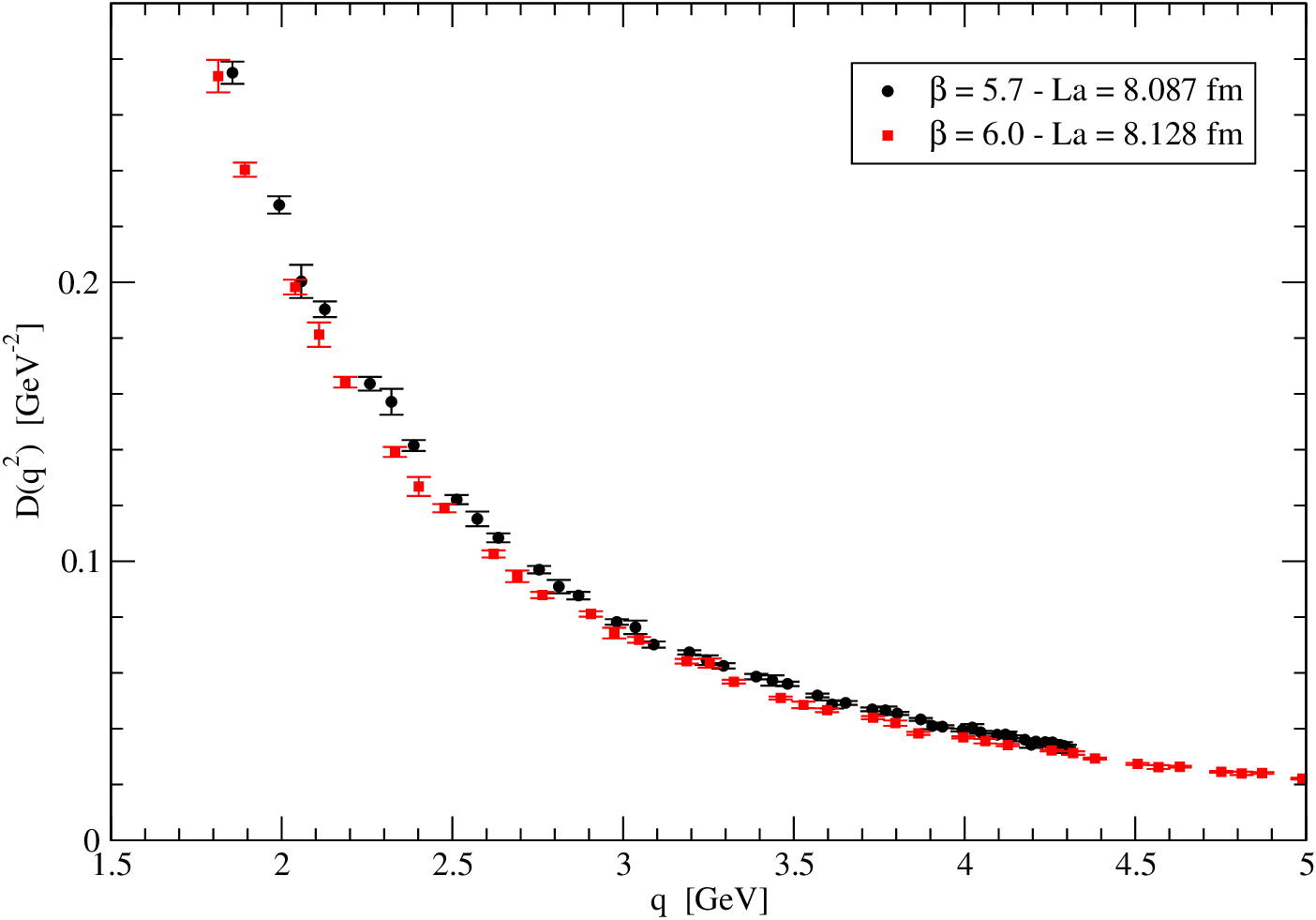} }\qquad
   \subfigure{ \includegraphics[scale=0.35]{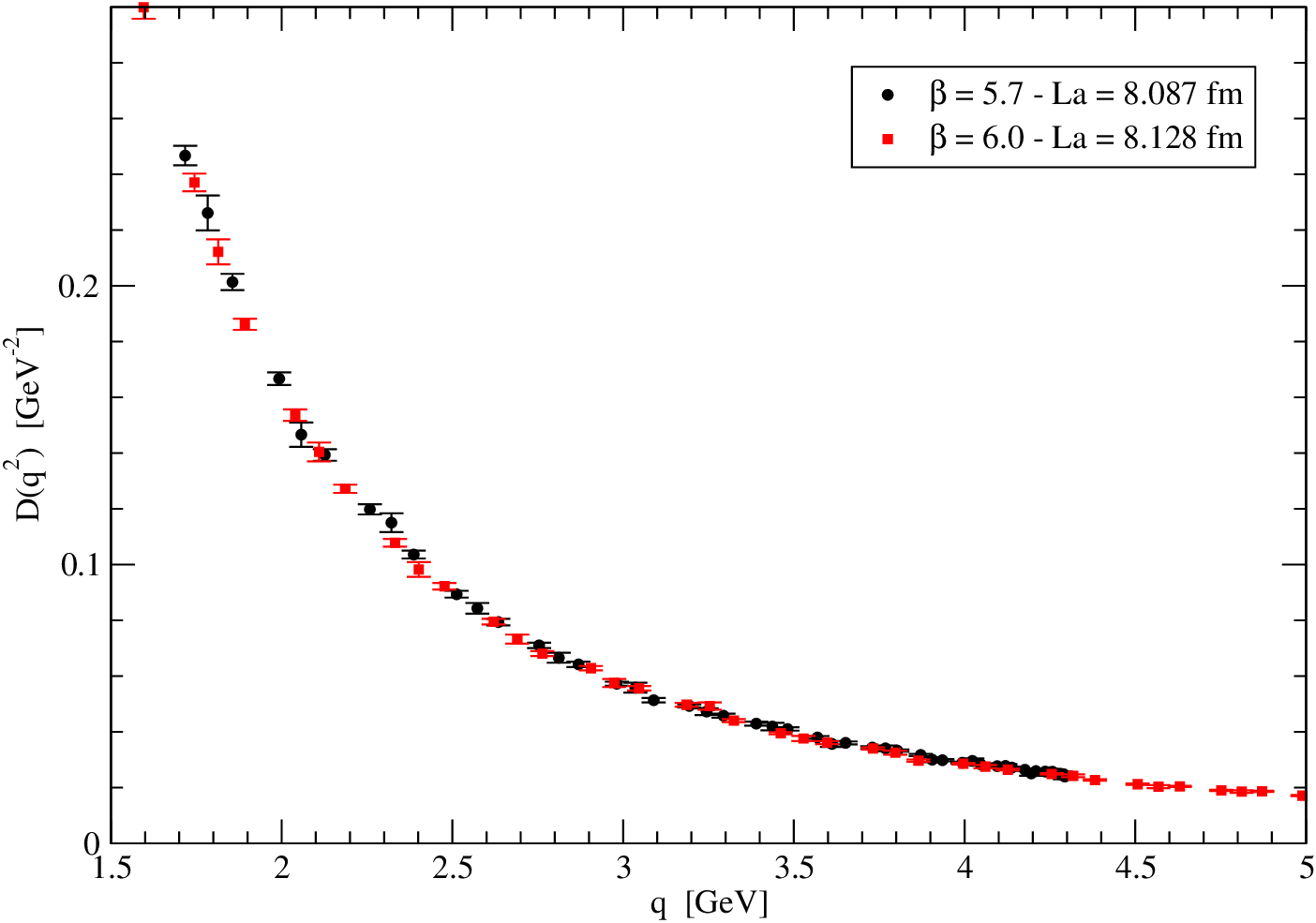}  }
  \caption{The gluon propagator renormalised at $\mu = 500$ MeV (left) and at $\mu = 1$ GeV  (right) 
                at intermediate momentum range (top) and up to 5 GeV (bottom).}
   \label{fig:glue1GeV500MeV_IR}
\end{figure*}

In Fig. \ref{fig:glue1GeV500MeV_IR} the details of the propagators renormalised at the different $\mu$
are shown for momenta up to 5 GeV. For $\mu = 1$ GeV, the difference between the two lattice computations
is washed out for momenta above $\sim 2$ GeV, with the data associated with the largest lattice spacing
($\beta = 5.7$) going always below the results computed with the smaller lattice spacing ($\beta = 6.0$).
This reproduces the behaviour observed in section \ref{spacing} ---
recall that for $\mu = 4$ GeV, the propagators agree within errors for momenta above $\sim 900$ MeV.
For $\mu = 500$ MeV, one can observe differences between
the computations of $D(q^2)$ both in the infrared, where the $\beta = 5.7$ data
goes below the $\beta = 6.0$ one,
and at intermediate and large momenta, where the $\beta = 5.7$ propagator goes above the $\beta = 6.0$ 
data. The two computations seem to become closer as $q$ approaches the ultraviolet but, within the range of
momenta accessed by the simulation, the two data sets hardly agree when one choses $\mu = 500$ MeV. In this
case the simulation shows differences between the two calculations in the full momentum range, i.e. both in
the infrared region and in the intermediate momenta.

\end{appendix}

According to what is observed in Figs. 
\ref{fig:glue1GeV500MeV} and
\ref{fig:glue1GeV500MeV_IR}, we conclude that the lattice artifacts due to a finite 
lattice spacing play an important role in the 
determination of the infrared propagator. Moreover, our analysis 
points towards a dominant effect in
the infrared propagator associated with the finite lattice spacing 
and not with the physical volume of the
simulation. In particular, it has been shown that a large lattice spacing such as $a \sim 0.18$ fm tend to suppress $D(q^2)$ at 
low momenta. Curiously, the value of
$D(0)$ does not seem to be so sensible to $a$. However, such a behaviour can be due to a larger statistical error on 
the computation of $D(0)$ due to the impossibility of performing a $Z_4$ average.


\end{document}